\documentclass{article}

\usepackage{arxiv}

\usepackage[utf8]{inputenc} % allow utf-8 input
\usepackage[T1]{fontenc}    % use 8-bit T1 fonts
\usepackage{hyperref}       % hyperlinks
\usepackage{url}            % simple URL typesetting
\usepackage{booktabs}       % professional-quality tables
\usepackage{amsfonts}       % blackboard math symbols
\usepackage{nicefrac}       % compact symbols for 1/2, etc.
\usepackage{microtype}      % microtypography
\usepackage{lipsum}
\usepackage{graphicx}
\title{Fast transient spray cooling of a hot thick target}

\author{
Fabian M. Tenzer, Ilia V. Roisman,  Cameron Tropea\\
  Institute for Fluid Mechanics and Aerodynamics, Technische Universit\"at Darmstadt\\
Alarich-Wei\ss-Stra\ss e 10, 64287 Darmstadt, Germany\\
  \texttt{roisman@sla.tu-darmstadt.de} \\
}

\begin{document}
\maketitle

\begin{abstract}
Spray cooling a hot target is characterized by strong heat flux density and fast change of the temperature of the wall interface. The heat flux density during spray cooling is determined by the instantaneous substrate temperature, which is illustrated by boiling curves. The variation of the heat flux density is especially notable during different thermodynamic regimes: film, transitional and nucleate boiling.

In this study transient boiling curves are obtained by measurement of the local and instantaneous heat flux density produced by sprays of variable mass flux, drop diameter and impact velocity. These spray parameters are accurately characterized using a phase Doppler instrument and a patternator. The hydrodynamic phenomena of spray impact during various thermodynamic regimes are observed using a high-speed video system.

A theoretical model has been developed for heat conduction in the thin expanding thermal boundary layer in the substrate. The theory is able to predict the evolution of the target temperature in time in the film boiling regime. Moreover, a remote asymptotic solution for a heat flux density during the fully developed nucleate boiling regime is developed. The theoretical predictions agree very well with the experimental data for a wide range of impact parameters.
\end{abstract}

% keywords can be removed

\section{Introduction}
Spray cooling is a process capable of achieving very high, nearly uniform heat flux densities and therefore high cooling performance. Spray cooling is used in various industrial applications, like cooling of  micro-chips and other high powered electronics or electrical parts \cite{mudawar2001,Bar-Cohen2006,Ebadian2011}, cooling of metal products in metallurgy during quenching processes, in  metalworking   \cite{Chen1992}, cooling of tools for hot forging \cite{Pola2013}, of solar panels \cite{Nizetic2016,Sargunanathan2016} and in many other technological processes.

The phenomena of spray impingement onto a very hot substrate can be significantly influenced by the wall temperature since the flow generated by each drop impact is influenced by various micro-scale thermodynamic effects, governed by an intensive evaporation. The hydrodynamics and heat transfer during single drop impact onto a heated wall have been intensively investigated   \cite{Chandra1991,Bernardin1997b,Bertola2015a,Staat2015a}. The regimes of single drop impact %\cite{Bernardin1997b,Bertola2015a,Staat2015a}
observed in the experiments
include single phase cooling, nucleate boiling, transition regime, thermal atomization \cite{Roisman2018} and film boiling, if the wall temperature rises above the Leidenfrost condition.

%A comprehensive review of the involved phenomena and existing modelling approaches can be found in \cite{Liang2017c,Breitenbach2018}.

A detailed review of the current state of the art in the field of spray cooling can be found in \cite{Liang2017b,Liang2017d,Cheng2016a,Kim2007,Breitenbach2018}. Many studies deal with the influence of different parameters on the performance of spray cooling. These studies are mostly focused on the determination of the critical (maximum) heat flux and on  obtaining the boiling curves, which describe the dependence of the heat flux density on the substrate temperature.  An example for empirical correlations can be found in \cite{Mudawar1994}. Among the governing parameters are spray properties, like droplet velocity, droplet diameter, mass flux density, or liquid properties \cite{Puschmann2004a,Wendelstorf2008,Yang1996,Chen2002a,Estes1995a}.  Moreover, experiments of \cite{Cebo-Rudnicka2016} with different target materials demonstrated that the heat transfer is influenced also by the  thermal conductivity of the surface.

Most of the models for the heat flux density and for the critical heat flux are completely empirical. The main goal of the present study is to develop a predictive theoretical model for fast transient cooling of a very hot thick substrate by spray impact based on the identification of the main influencing physical parameters.  These influencing parameters are different for the film boiling regime and for the nucleate boiling regime.

The model is developed using  measurements of heat flux density during impingement of an accurately characterized spray onto a initially heated  target, whose thickness is much thicker than the thickness of the thermal boundary layer which develops in the substrate.
The temperature and the heat flux measurements are accompanied by the  high-speed visualizations of an impacting spray at different time instants, which allows to identify various hydrodynamic and thermodynamic regimes of the spray cooling process.

A one-dimensional model for heat transfer associated with spray cooling  is developed which accounts for the development of the thermal boundary layer in the substrate. A remote asymptotic solution for the heat flux density is developed for the heat flux density in the fully developed nucleate boiling regime when the entire substrate surface is covered by a thin liquid film. The theoretical predictions the heat flux density agree very well with the experimental data.  The model accounts for different spray fluids, but it is only validated for water sprays having different fluid temperatures.

\section{Experimental methods}

\begin{figure}
	\centering\includegraphics[width=0.9\textwidth,keepaspectratio=true]{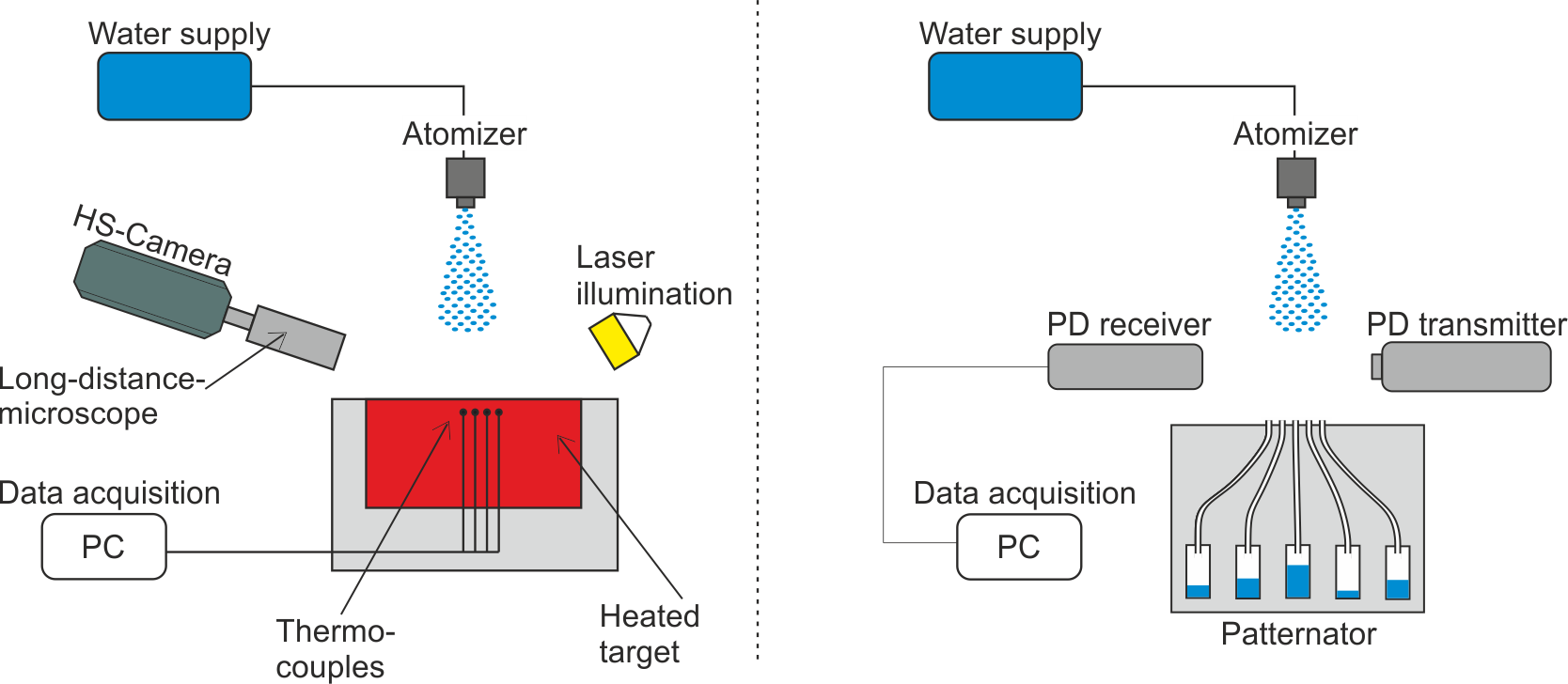}
	\caption{Schematic of experimental setup. Left: Heat flux measurements with thermocouples and visualization with HS-camera. Right: Spray characterization with phase Doppler measurement system and patternator.}
	\label{fig:setup}
\end{figure}

The experimental setup designed for spray cooling experiments consists of six main systems shown in Fig.~\ref{fig:setup}:  a water supply system, heated target, temperature measurement and control system, spray characterization system, optical observation system and data acquisition system.

A conventional pressure driven, full-cone atomizer is supplied with water, purified by a reverse osmosis device, from a pressurized tank.
%The spray system is located directly above the heated surface, since only normal spray impact is considered in this study.
By adjusting the distance between the nozzle and the heated surface, and varying the pressure supplied to the nozzle, sprays of different properties upon impact can be generated. The resulting spray is described by  the local properties: mass flux density $\dot{m}$,  mean drop diameter $D_{10}$ and  mean drop velocity $U$.
%A suction system is implemented to ventilate overspray, to ensure  optical access to the spray impact region, and to improve the reproducibility of the experimental results.

%\subsection{The substrate temperature and heat flux measurements}

The \textit{heated surface} of the spray impact target is the top end of a circular cylinder (diameter $d = 100\,\mathrm{mm}$ and height $h = 53.2\,\mathrm{mm}$) built of stainless steel (1.4841). The target is heated by four cartridge heaters with an overall power of $2\, \mathrm{kW}$.
All of the heaters are placed in a copper disc which is screwed to the bottom of the cylinder. The side and bottom of the target are insulated in order to assume that heat transfer only occurs through the top surface. The target is placed in a water resistant housing.
The average roughness of the impinging surface is $0.03\,\mathrm{\mu m}$. The setup is designed for surface temperatures up to $500\, \mathrm{^\circ C}$.

The \textit{temperature just below the substrate surface} is monitored using eight thermocouples (type K, class 1, isolated, $0.5\, \mathrm{mm}$ diameter) embedded inside the test specimen. Six of them are located in two rows (three each row) $0.5\, \mathrm{mm}$ and $3.5\, \mathrm{mm}$ below the impingement surface. Their radial spacing is $1.75\, \mathrm{mm}$ and the first is placed in the centre of the circular target.

 For control of the heating system two thermocouples are installed at the bottom and in the middle of the specimen. All thermocouples are placed inside holes with $0.6\, \mathrm{mm}$ diameter, which were produced using spark erosion. To ensure good thermal contact between the thermocouples and the material they are bonded inside the holes with a thermally high conductive adhesive (Aremco Ceramabond 569).

To evaluate the response time, a thermocouple was immersed from ambient air into water having a temperature of $80\, \mathrm{^\circ C}$ and the time until the temperature readings showed $90\, \%$ was measured. The quantified response time is in the range of $0.8\, \mathrm{s}$. This time is independent of the temperature increment.

The instantaneous radial temperature distribution in the target, the local heat flux density and the target surface temperature (where direct measurements are not possible) are computed by solving the two-dimensional inverse heat conduction problem, using the thermocouples as input data. The solution is obtained using an analytical approach and a code published by \cite{Woodfield2006}. The solution is based on solving the second order partial differential equation for heat conduction using a Laplace transform technique.

The \textit{observation system} consists of a  high-speed camera (Phantom v12.1 with the frame rate of $55,000\, \mathrm{fps}$) equipped with a Questar QM-100 long distance microscope and a pulsed Cavilux HF laser (pulse duration is $400\, \mathrm{ns}$) backlight used as an illumination source.
 %yields  images of high contrast. Both illumination and camera are inclined at approximately $20\, \mathrm{^\circ}$ to the horizontal target, resulting in an inclined line of sight, which prevents splashing droplets from obstructing the view onto the surface.
The field of view is approximately 3 mm in width at a resolution of $384 \times 240 \mathrm{\,px}$. The visualization and heat flux measurements are temporally matched and therefore visual observations can be directly associated with the instantaneous local heat flux density and target surface temperature.
 %At this resolution the camera is capable of recording at a frame rate of $55,000\, \mathrm{fps}$.  A high-speed pulsed laser (puls duration is $400\, \mathrm{ns}$) is used as the illumination source.
% At this repetition rate the laser is capable of providing a $400\, \mathrm{ns}$ illumination pulses for each camera frame, which is necessary to achieve adequate illumination in dense sprays, but is still short enough to avoid motion blur.

The \textit{main spray properties are accurately characterized} using a phase Doppler instrument (Dantec Dynamics, Dual - Mode) for measurements of the distributions of the drop diameter and two components of the drop velocity in the spray. Phase Doppler data is acquired under free stream conditions without the presence of the target. A custom built patternator is used for measurement of the spatial distribution of the local mass flux density of the spray.
 The patternator collects the spray in  17 small tubes of  inner diameter of $4\, \mathrm{mm}$  placed  in a row with a distance of $6\, \mathrm{mm}$ to one another. Similar devices are described in \cite{Lefebvre2018}. The orifices of the tubes are positioned at the top surface of a bluff body having the same dimensions as the heated target to ensure the same airflow encountered during cooling experiments. The high spatial resolution resolves inhomogeneities of the spray, providing the local mass flux density at the position where the heat flux measurements have been computed.

\begin{figure}
	\centering\includegraphics[width=0.49 \textwidth]{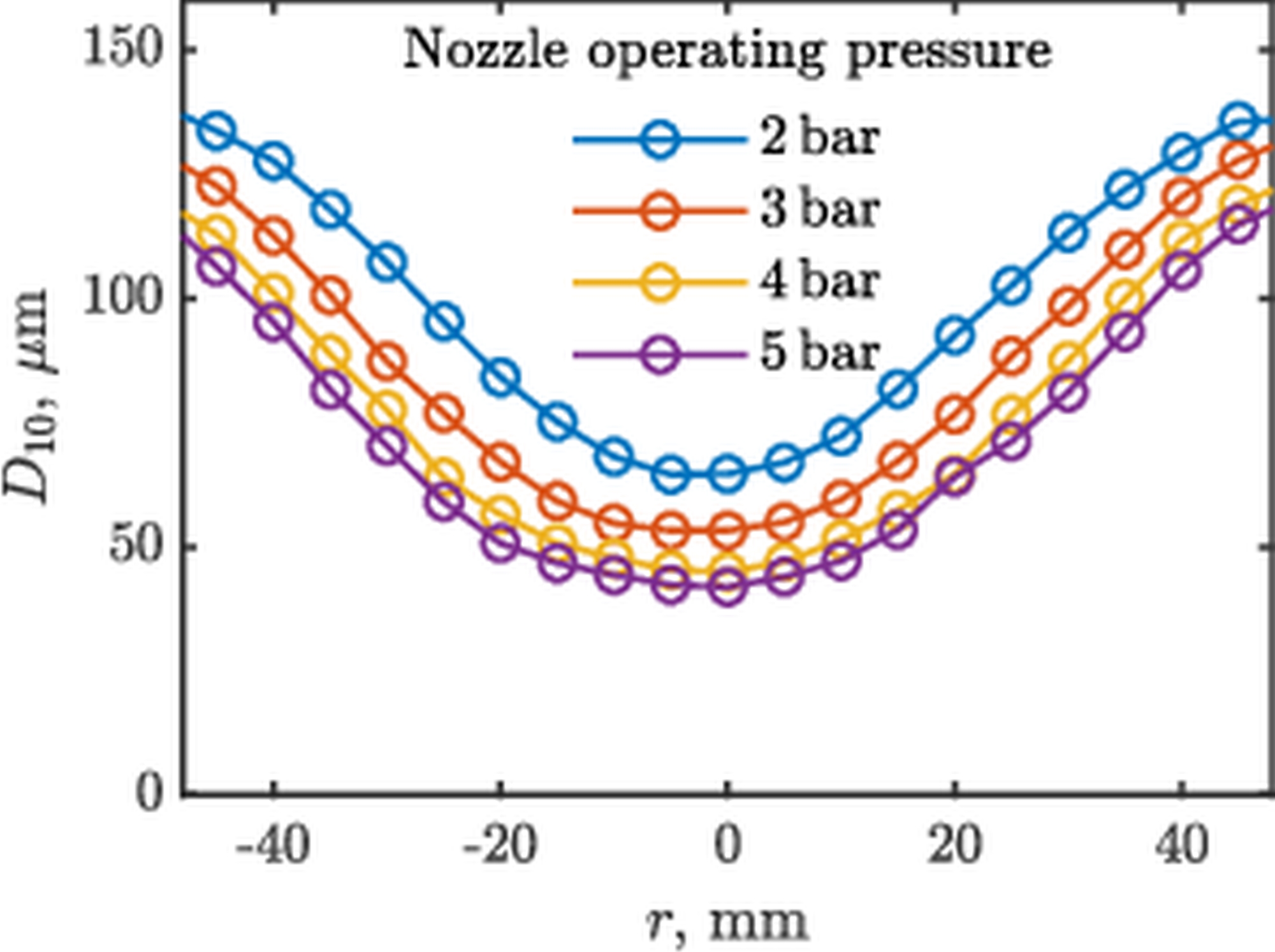}\includegraphics[width=0.49 \textwidth]{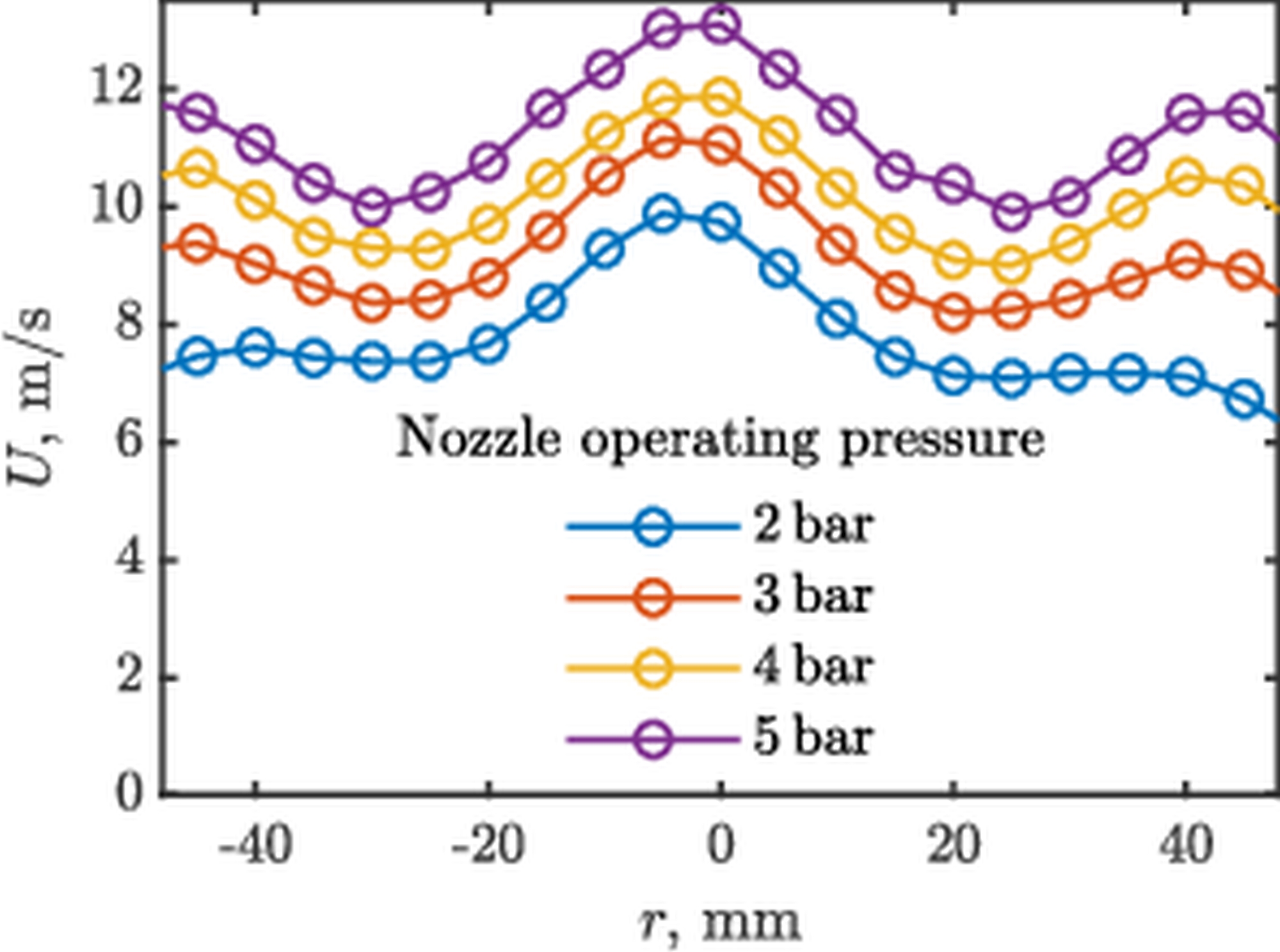}\\
	\vspace{0.5 cm}
	\includegraphics[width=0.49 \textwidth]{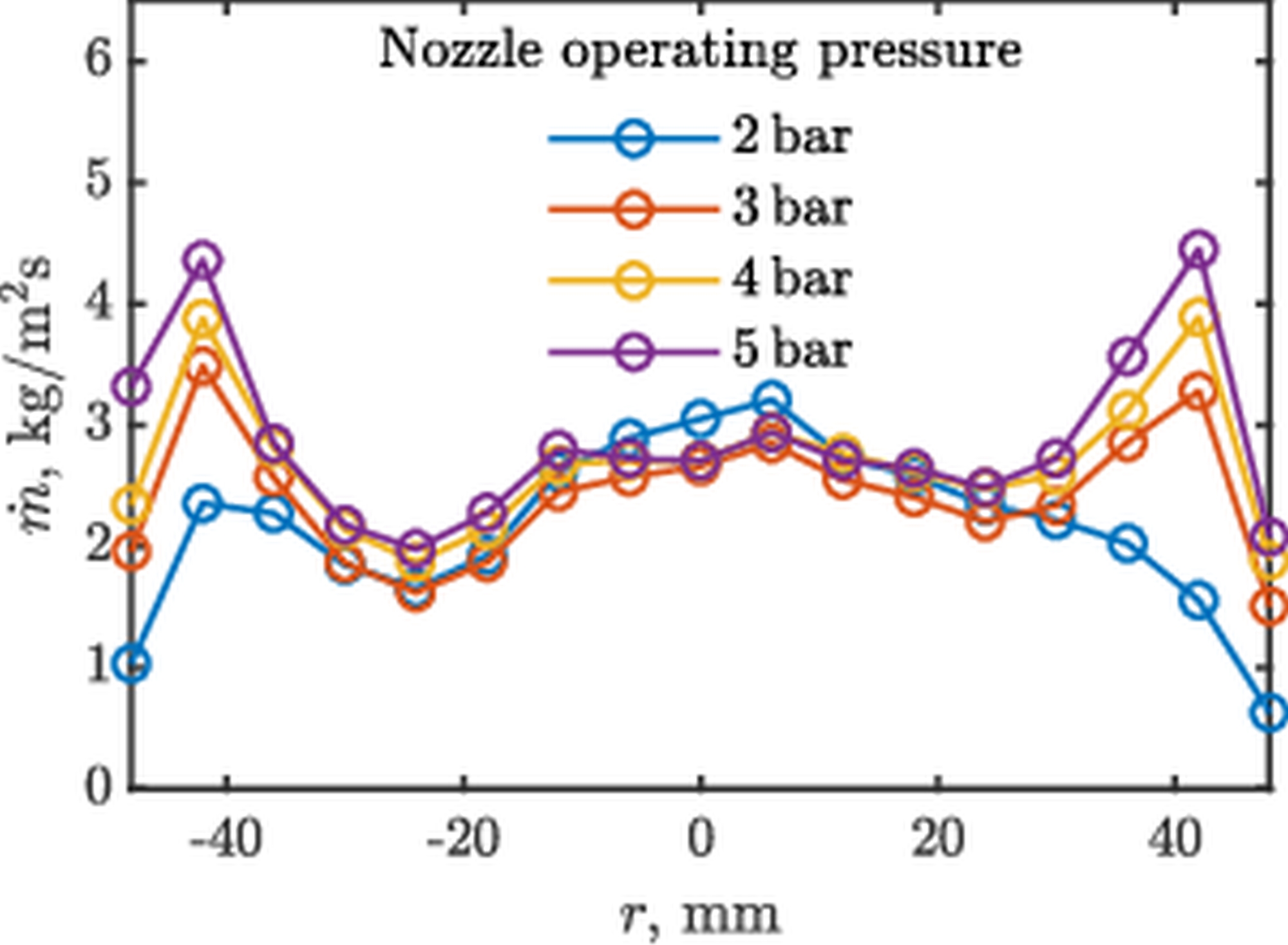}
	\caption{ Mean drop diameter ($D_{10}$), mean axial velocity of drops ($U$) in the impacting spray and the axial mass flux $\dot m$ as a function of radial position. Position $r=0\, \mathrm{mm}$ corresponds to the  axis of the nozzle orifice.}
	\label{fig:D}
\end{figure}

In Fig.~\ref{fig:D}  the mean drop diameter $D_{10}$, the mean impact velocity $U$ and the local mass flux density $\dot m$  are  shown at $100\, \mathrm{mm}$  from the nozzle as a function of the radial coordinate $r$. The position $r=0$ corresponds to the axis of the nozzle. As expected, the  average drop diameter reaches a minimum at the axis and decreases for higher injection pressures. The average drop velocity is maximum at the axis.

It is interesting that the mass flux density at the center depends only  weakly  on the injection pressure. For the nozzles used in the experiments the pressure influences mainly the flux distribution in the outer ring of the spray cross-section. In these experiments the magnitude of the mass flux density is varied by changing the distance between the nozzle and the cooling target and the injection pressure.

The gradients of the main spray integral properties near the center are rather small. The spray near the axis on a spot of radius $5 - 10\, \mathrm{mm}$ can be considered as nearly uniform.

{Since the phase Doppler measurements were performed in absence of the target, the question arises whether these free stream results are comparable to those in presence of the heated target. For that reason we performed additional measurements in presence of a bluff body having the same dimensions as the heated target. The measurement plane was located $10\, \mathrm{mm}$ above the surface of the target. During data processing we included only those droplets having a mean velocity in the downward direction,  to exclude those droplets that had already impacted at the surface and rebounded. In an exemplary case the integral spray characteristics are for free stream conditions: $D_{10}=59\, \mathrm{\mu m}$ and $U=9.6\, \mathrm{m/s}$. Due to the presence of the bluff body the values changed to: $D_{10}=81\, \mathrm{\mu m}$ and $U=9.7\, \mathrm{m/s}$. Especially the small droplets are affected by the displacement of the target resulting in the higher $D_{10}$, since only a smaller portion of them reach the surface. However, because the largest influence on the heat flux density is the mass flux density and the previously mentioned differences are rather small, we conclude that it is reasonable to use the droplet diameter and velocity acquired under free stream conditions as input data for the present work. Perhaps even more significant is however the fact that the measurement of the mass flux density using the patternator does mimic exactly the conditions prevailing with the heated target.

\section{Measurements of the heat flux density}
\begin{figure}
	\centering\includegraphics[width=.5\textwidth]{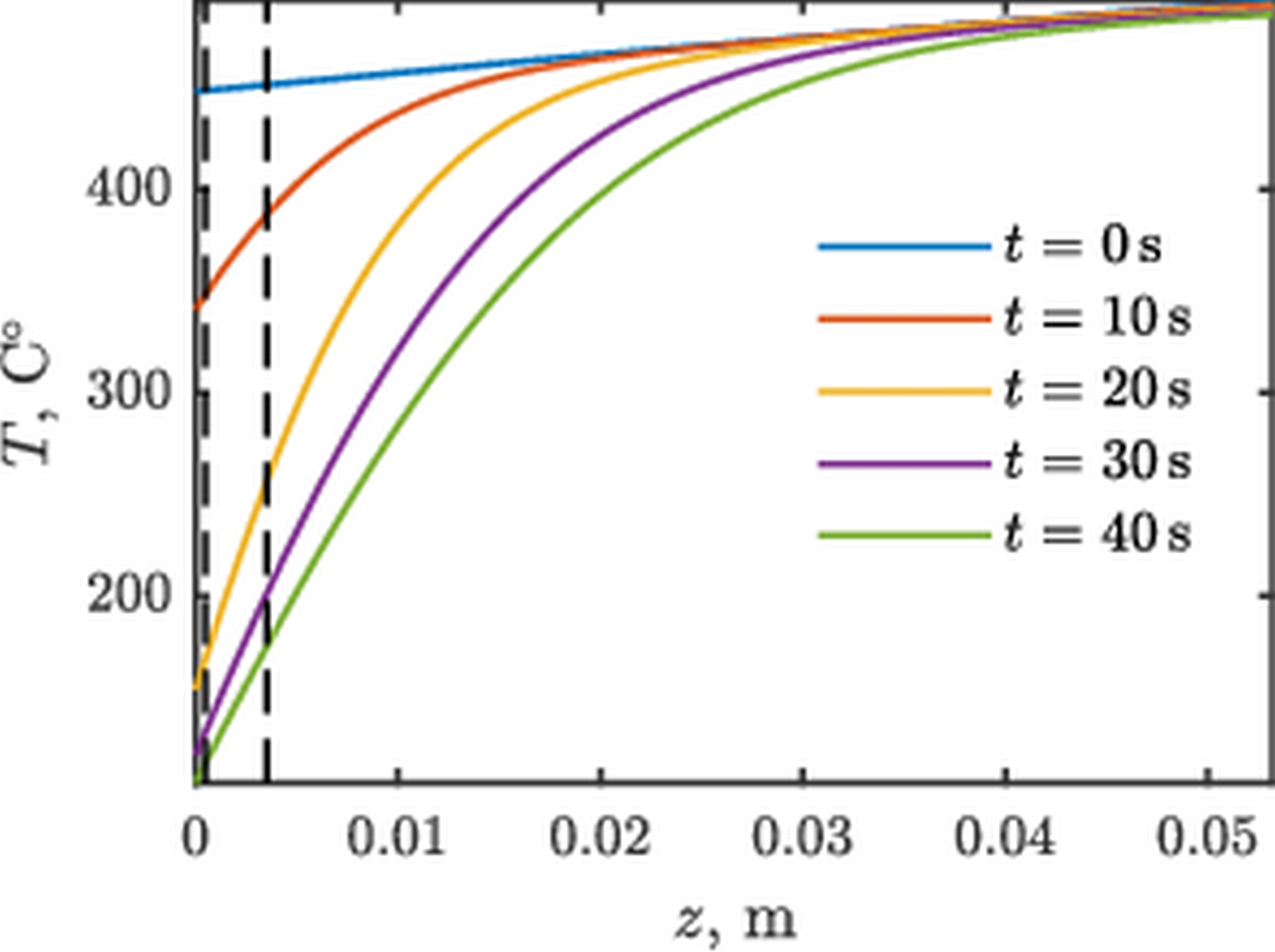}
	\caption{ Calculated temperature $T$ inside the substrate as a function of the depth $z$ for different times $t$. The dashed lines indicate the position of the the thermocouples at $z_1=0.5\, \mathrm{mm}$ and $z_2=3.5\, \mathrm{mm}$.}
	\label{fig:gradient}
\end{figure}

In many studies of heat transfer, in order to evaluate the boiling curve, the setup is designed to keep the substrate temperature constant. In this study the effect of the transient phenomena is investigated, since this situation is relevant to many industrial applications mentioned in the introduction section.

Therefore, the target is initially heated until it achieves a given practically uniform initial temperature. The spray parameters are kept constant during the entire experiment. The influence of the start-up phase during the initial  development of the spray is avoided by using a shutter in front of the nozzle.The experiment is started only when the spray is fully developed and the target is heated uniformly. At the instant $t=0$ the  heating of the target is switched off simultaneously with the opening of the spray shutter. At this instant the substrate temperature starts to change due to the heat flux associated with spray impact.

The typical evolution of the temperature field inside the substrate, the surface temperature and the heat flux density are illustrated in Figs.~\ref{fig:gradient}, \ref{fig:time_series} and \ref{fig:q_T_450}. The initial wall temperature is $450\, \mathrm{^\circ C}$. The spray parameters for this case are: $\dot{m}=2.9\, \mathrm{kg/m^2s}$, $D_{10}=55\, \mathrm{\mu m}$ and $U=10.3\, \mathrm{m/s}$.

The evolution of the temperature profiles $T(z)$, calculated by solving the inverse heat conduction problem, are shown in Fig.~\ref{fig:gradient} for different times $t$.  The $z$ coordinate coincides with the spray axis while the position $z= 0$ corresponds to the wall surface, where the spray impact takes place. The bottom of the heated plate corresponds to $z= 53.2\, \mathrm{mm}$.

The vertical dashed lines in Fig.~\ref{fig:gradient} indicate the position of the the thermocouples at $z_1=0.5\, \mathrm{mm}$ and $z_2=3.5\, \mathrm{mm}$. The temperature measurements of the thermocouples are used as the input data in the solution of the inverse heat conduction problem.

The corresponding time series of the surface temperature and heat flux density are plotted in Fig.~\ref{fig:time_series}. The vertical dashed lines in Fig.~\ref{fig:time_series} correspond to the bounds of the boiling regimes: film boiling, transition boiling and nucleate boiling regime, which are described in more detail in Figs.~\ref{fig:q_T_450} and \ref{fig:q_bilder}.

\begin{figure}
	\centering\includegraphics[width=0.49 \textwidth]{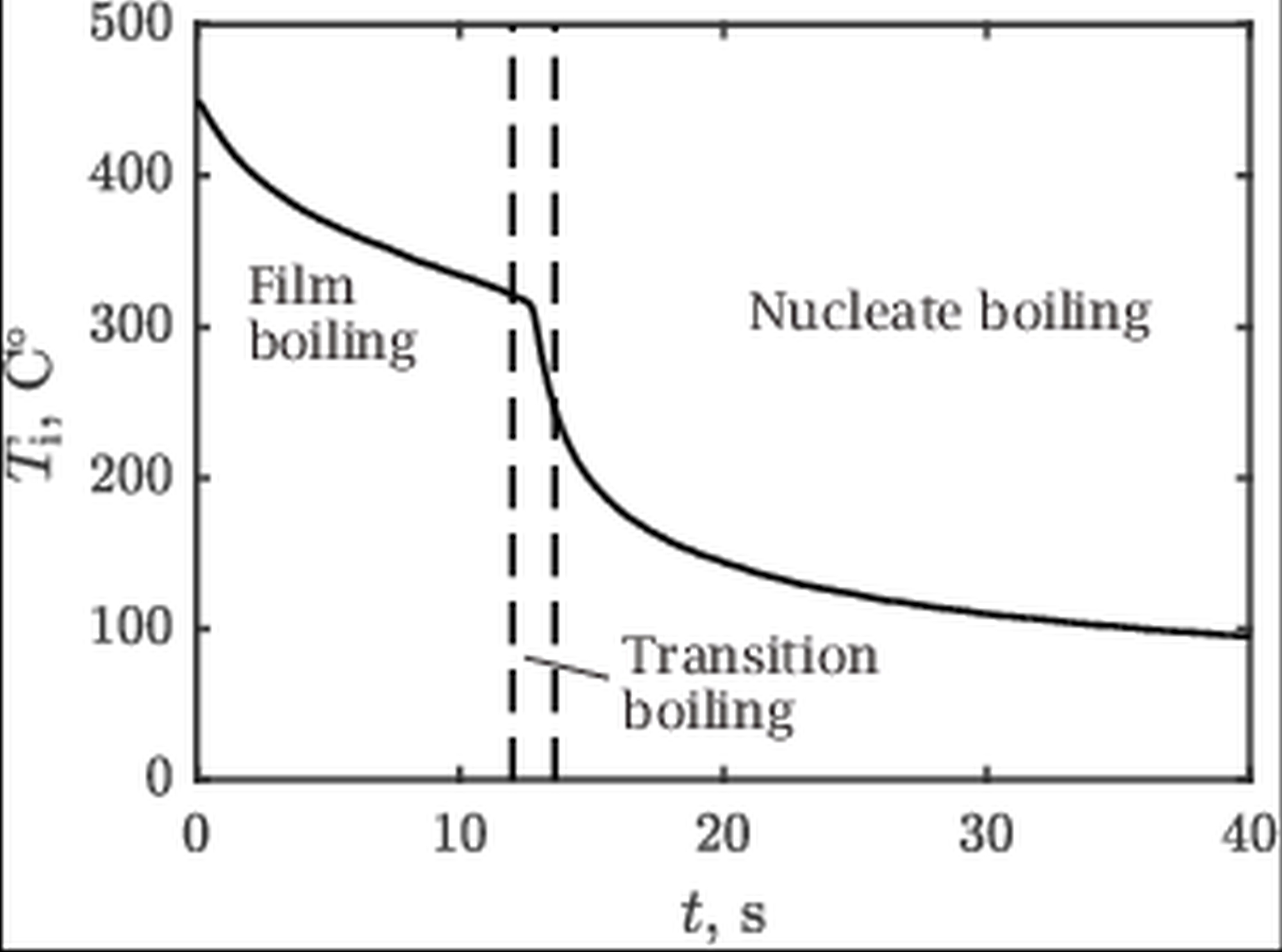}\includegraphics[width=0.49 \textwidth]{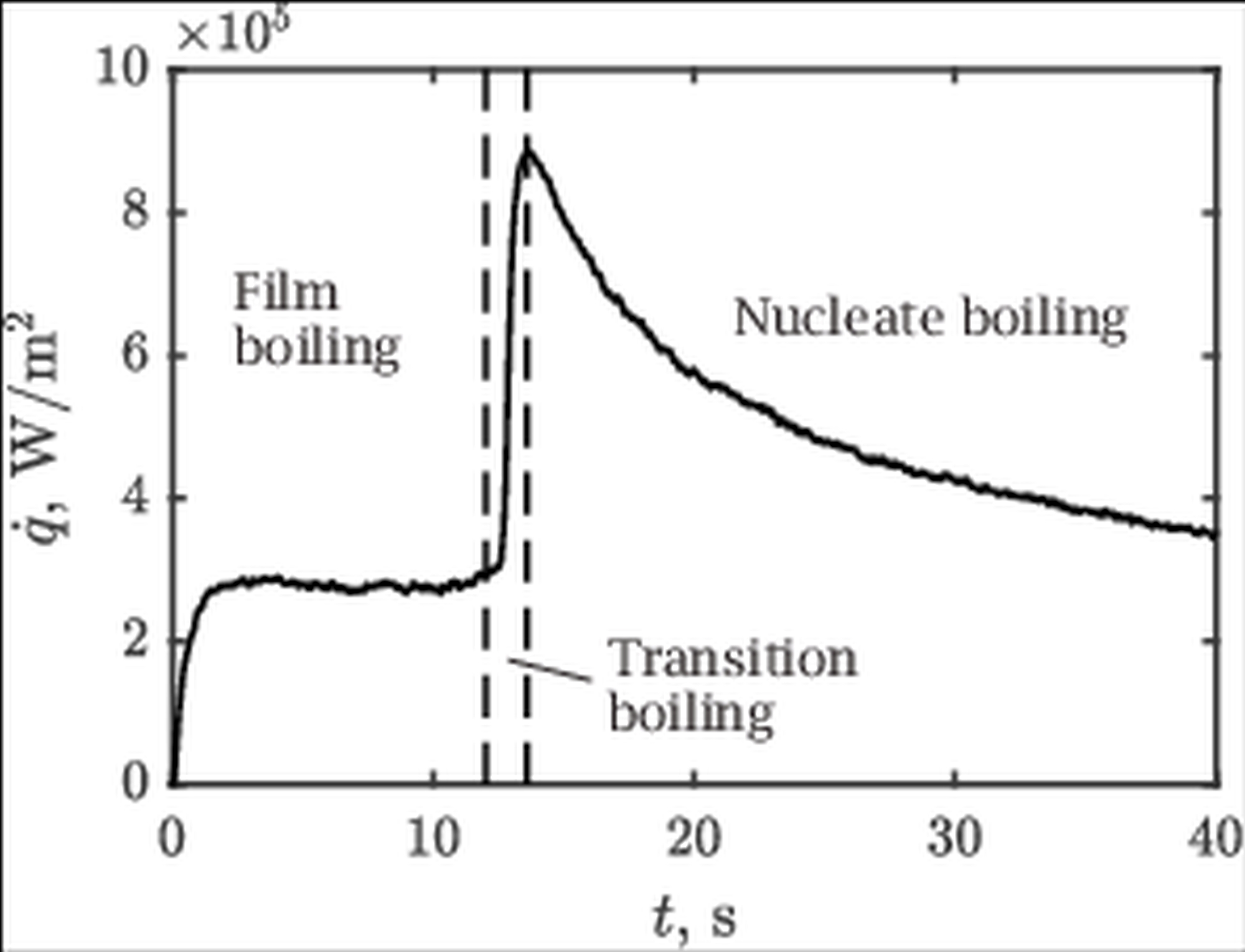}\\
    \caption{Typical temporal evolution of the  surface temperature $T_\mathrm{i}(t)$ and instantaneous local heat flux density $\dot{q}(t)$. The vertical dashed lines indicate the boundaries between the film boiling, transition and nucleate boiling regimes.}
	\label{fig:time_series}
\end{figure}

\begin{figure}
	\centering\includegraphics[width=\textwidth]{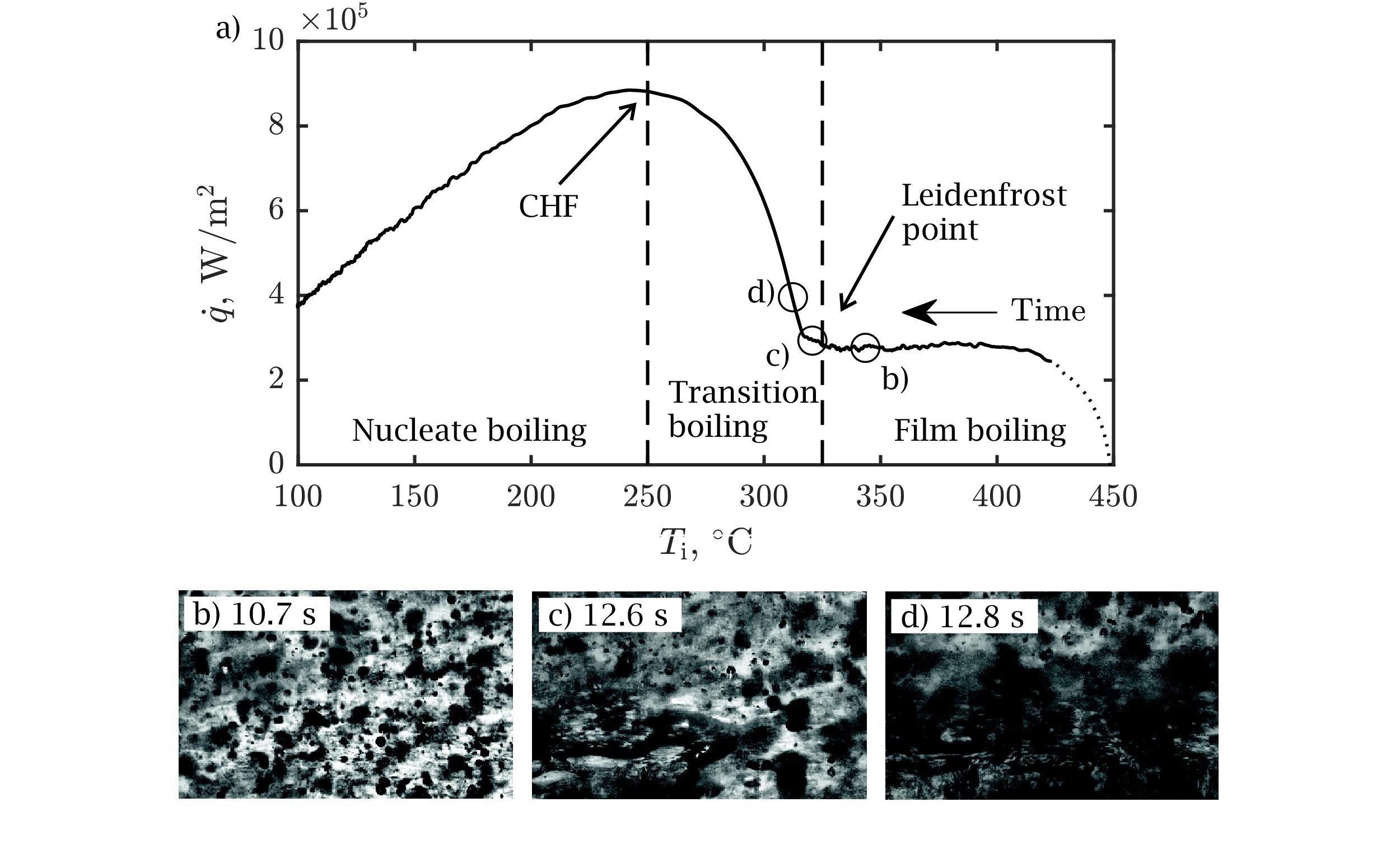}
	\caption{Spray cooling regimes at different surface temperatures $T_\mathrm{i}(t)$ around the Leidenfrost point. a) Measured heat flux density $\dot q$ as a function of surface temperature $T_\mathrm{i}$; b) visualized spray impact in the film boiling regime; c) at the Leidenfrost point, characterized by the first appearance of liquid patches; and  d) the fast expansion of the liquid spots and increased influence of nucleate boiling. The videos of the spray impact at different time instants are provided in the supplementary material.}
	\label{fig:q_T_450}
\end{figure}

In Fig.~\ref{fig:q_T_450} the instantaneous local heat flux density $\dot{q}(t)$ in the central area of the target and spray is  plotted as a function of the surface temperature $T_\mathrm{i}(t)$ for the same experimental data, whereby time $t$ increases following the curve to the left.

 The precision of the thermal measurements at the initial stage of cooling is not high and this  is caused by the very high temperature gradients during the initial fast temperature changes. The rise time of the thermocouples is not short enough and the thermal inertia of the material between the tip of the thermocouple and the surface is too high to precisely capture these fast temperature changes. This part of the plot, where the measurement precision is not well quantified, is indicated  here and on the following figures by a dotted curve.

In  Figs.~\ref{fig:q_T_450} b), c) and d)  images of spray impact and hydrodynamic phenomena at the surface captured at different instants after the spray cooling begins are shown. The time instants chosen for these images are marked on the graph in Fig.~\ref{fig:q_T_450} a).  The corresponding movie of spray impact is provided in the supplementary material.

The hydrodynamic phenomena visualized in  Figs.~\ref{fig:q_T_450} b), c), and d) are each different, since they correspond to different drop and spray impact thermodynamic regimes.

 In Fig.~\ref{fig:q_T_450} b) each drop impact onto the wall leads to its break up, formation of multiple  secondary droplets \cite{Roisman2018} and rebound. The contact time is short and there is no remaining wetting of the surface. As a result, the heat flux density is low, which is typical for the film boiling regime.

 At the Leidenfrost point, illustrated in Fig.~\ref{fig:q_T_450} c), a few  impacting drops start to wet and spread on the surface - a part of the surface is covered by initial liquid patches.

\begin{figure}
	\centering\includegraphics[width=\textwidth]{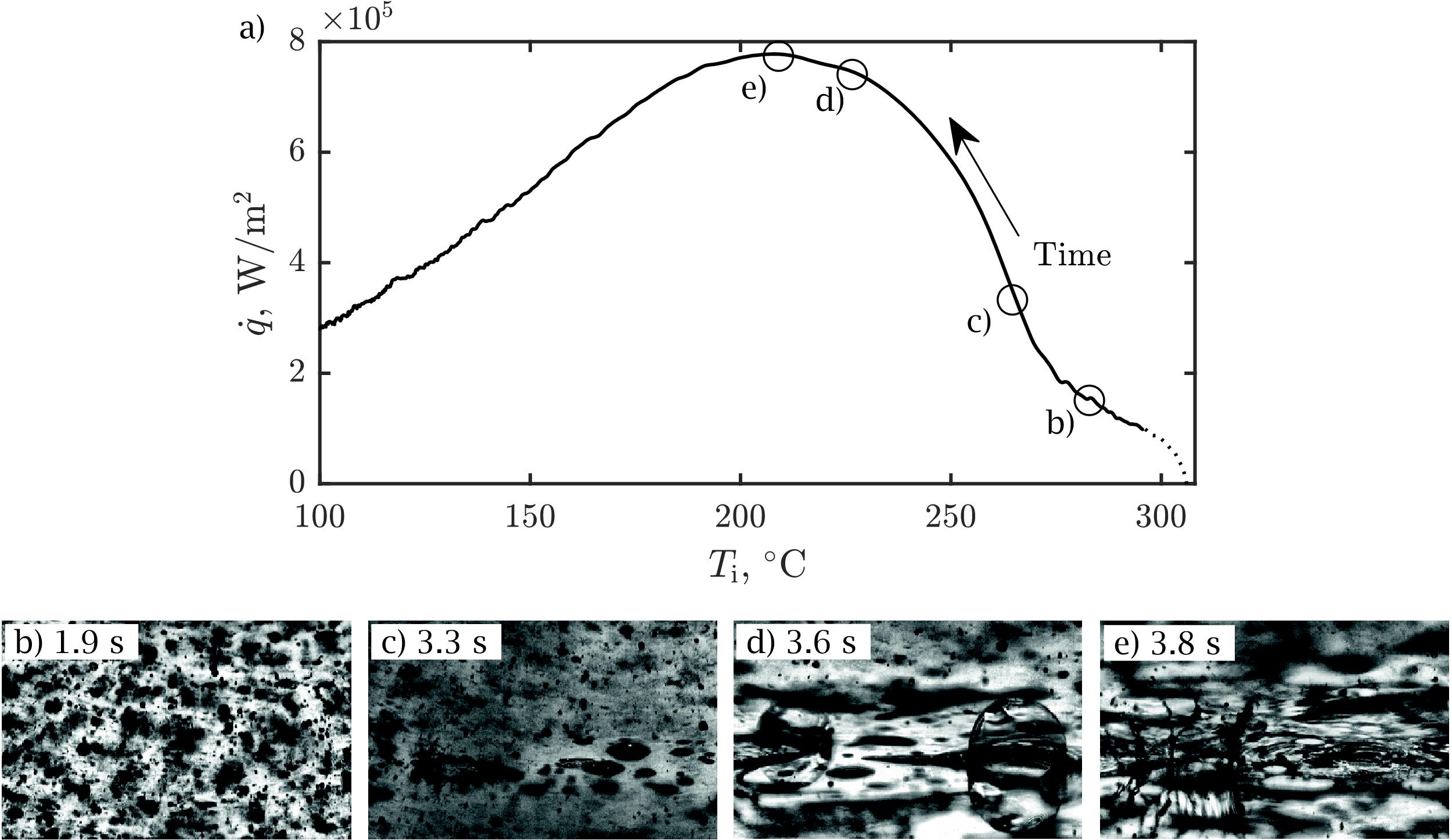}
	\caption{Phenomena of spray impact regimes at different surface temperatures $T_\mathrm{i}(t)$. a) Measured heat flux density $\dot q$ as a function of surface temperature $T_\mathrm{i}$; b) Image of the substrate exposed to spray impact in the film boiling regime; c) inception of the transition regime d) image corresponding to the fast expansion of the wetted area; e) apparently completely wetted surface at the instant corresponding to the critical heat flux. The videos of the spray impact at different time instants are provided in the supplementary material.}
	\label{fig:q_bilder}
\end{figure}

 At the next instant the area of the wet patches increases, Fig.~\ref{fig:q_T_450} d), and the heat flux density starts to rapidly increase. This phenomena correspond to the transition boiling regime. }

 Similar phenomena in the film and transition regimes are observed in Figs.~\ref{fig:q_bilder} b) and c), respectively. The images are of higher contrast, since the mass flux of the spray is smaller in the experiment shown in Fig.~\ref{fig:q_bilder}. In Fig.~\ref{fig:q_bilder} a) the measured heat flux density is plotted as a function of the surface temperature measured during continuous spraying. In the illustrated case the spray properties are: $\dot{m}=0.9\, \mathrm{kg/m^2s}$, $D_{10}=43\, \mathrm{\mu m}$ and $U=9.8\, \mathrm{m/s}$. In this example the target is initially heated to a surface temperature of $306\, \mathrm{^\circ C}$.

 Shortly before the point where the maximum heat flux is achieved a large portion of the surface is wetted by a liquid water film, as shown in Fig.~\ref{fig:q_bilder} c). Small nucleation bubbles form, grow and collapse. The heat flux density increases significantly, since the wetted area of the substrate  increases rapidly. Heat goes into the overheating of the liquid (sensible heat) and into the formation of bubbles.

 At the instant corresponding to the critical heat flux, Fig.~\ref{fig:q_bilder} d) $170\, \mathrm{^\circ C}$, the surface area is almost completely wetted by liquid. The appearance of a corona of an impacting  drop is clear evidence that the drop impacts onto a liquid film.

 At larger times the substrate is completely covered by a thin boiling liquid film. The film is continuously fed by fresh water from the spray and increases in coverage and depth. The heat flux density   reduces with time. This regime corresponds to fully developed nucleate boiling of spray cooling.

\begin{figure}
	\centering\includegraphics[width=.45\textwidth]{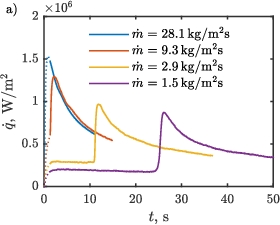}\hspace{0.5 cm}\includegraphics[width=.45\textwidth]{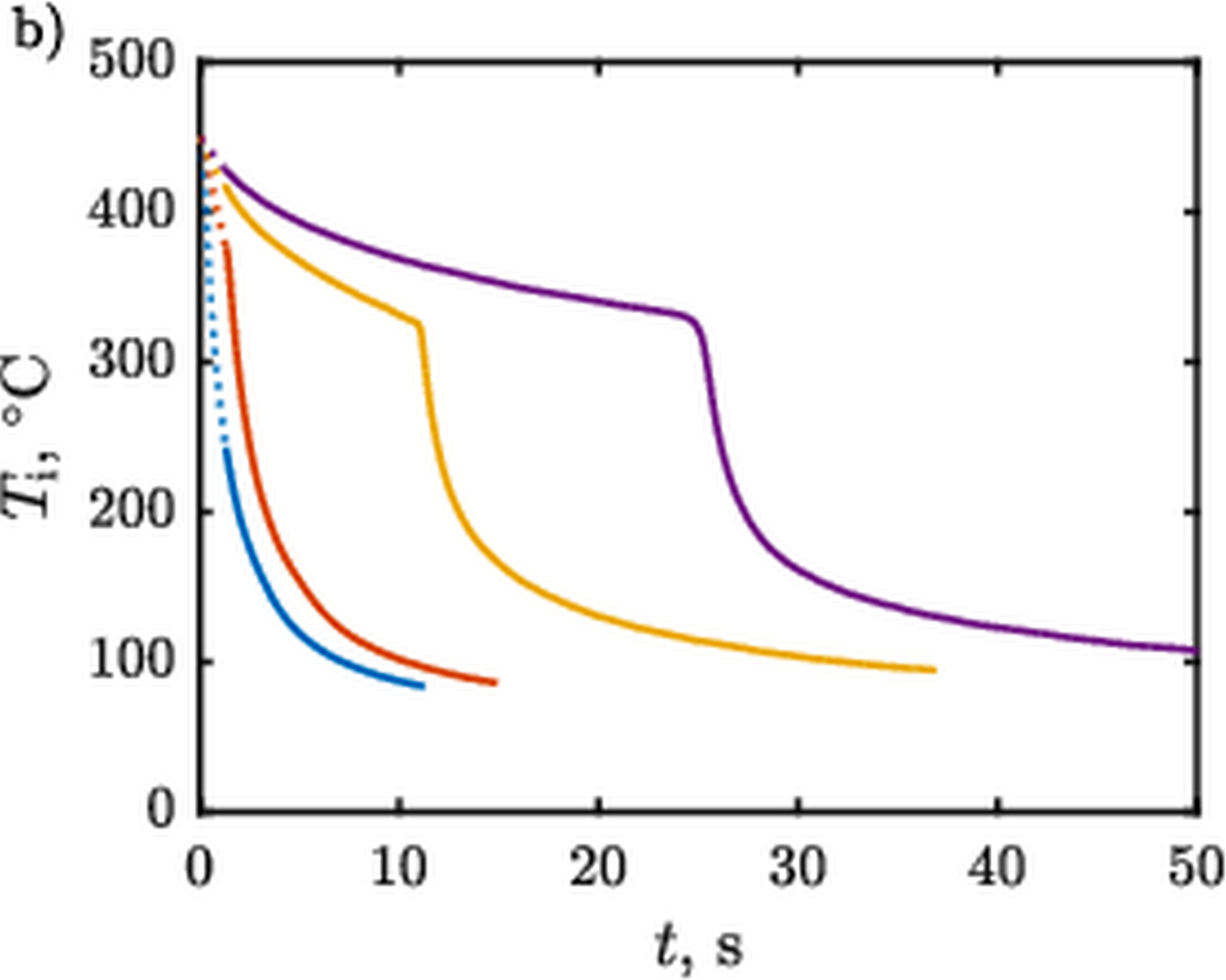}
	\caption{ Influence of different mass flux densities $\dot{m}$ on: a) Heat flux density and b) surface temperature  $T_\mathrm{i}$ dependence on time $t$.}
	\label{fig:q_m}
\end{figure}

 To better understand the influence of the main spray parameters on heat transfer, measurements with different mass flux densities were performed. Fig.~\ref{fig:q_m} a) shows the heat flux density $\dot{q}$ as a function of time $t$ for the mass fluxes $28.1$, $9.3$, $2.9$ and  $1.5\, \mathrm{kg/m^2s}$. The variation of other spray parameters remain in a relatively narrower range ($D_{10}=43-52\, \mathrm{\mu m}$ and $U=11.2-17.4\, \mathrm{m/s}$). The initial substrate temperature is $450\, \mathrm{^\circ C}$ for all the tests in Fig.~\ref{fig:q_m}. The experiments were stopped when the first thermocouple reading reached $100\, \mathrm{^\circ C}$. Increasing the mass flux density results in an increased heat flux density at all times and boiling regimes. The corresponding surface temperature $T_\mathrm{i}$ as a function of time $t$ is shown in Fig.~\ref{fig:q_m} b).

%\begin{figure}
%	\centering\includegraphics[width=.45\textwidth]{q_T_infl_m.jpg}\hspace{0.5 cm}\includegraphics[width=.45\textwidth]{q_T_infl_init.jpg}
%	\caption{Heat flux density dependence on the surface temperature: a) for different mass flux densities. The initial substrate temperature is $450\, \mathrm{^\circ C}$ for all the tests; and b) for different initial substrate temperatures.}
%	\label{fig:q_m}
%\end{figure}

%\begin{figure}
%	\centering\includegraphics[width=.45\textwidth]{q_T_infl_m_t.jpg}\hspace{0.5 cm}\includegraphics[width=.45\textwidth]{q_T_infl_init_t.jpg}
%	\caption{\textbf{Gleicher Inhalt wie  Fig. \ref{fig:q_m}. Welche Figure veroeffentlicht werden soll, muss noch besprochen werden.}}
%\end{figure}

The slopes of the curves are similar, especially in the nucleate boiling regime, and differ mainly at high temperatures between the start of the cooling experiments and the Leidenfrost point. In this region the heat flux density is much lower for sparse sprays ($1.5$ and $2.9\, \mathrm{kg/m^2s}$) than for the more dense sprays ($9.3$ and $28.1\, \mathrm{kg/m^2s}$). Moreover, for $\dot{m}=9.3$ and $28.1\, \mathrm{kg/m^2s}$ no film boiling regime can be identified, since the time of the film boiling regime is very short. This can be explained by the limited response time of the thermocouples which prevents the detection of the Leidenfrost point and film boiling regime in the case of this very fast cooling process.

%In Fig.~\ref{fig:q_m} images of the surface having the same surface temperature of $320\, \mathrm{^\circ C}$ are shown for different mass flux densities. For $0.9\, \mathrm{kg/(m^2s)}$ only a small area of the target is covered by liquid. There are only few connected ligaments but more sessile droplets. The corresponding heat flux density is $0.3\, \mathrm{MW/m^2}$. At $2.8\, \mathrm{(kg/m^2s)}$ the surface is covered mainly by a connected liquid film, resulting in a larger wetted surface area where heat transfer can occur. The heat flux density in this regime slightly increases to $0.35\, \mathrm{MW/m^2}$. Further increasing the mass flux causes the surface to be completely covered with liquid and yields to a strong rise in heat flux density to $1.3\, \mathrm{MW/m^2}$. Although the surface is already nearly completely covered by a liquid film for $2.8\, \mathrm{(kg/m^2s)}$, another increase of the mass flux density again increases the heat flux. This can be explained by the droplets splashing onto the heated film and thereby causing nucleation bubbles to detach from the surface and yield to fresh cold liquid. As $0.9\, \mathrm{(kg/m^2s)}$ represents a very sparse spray, it becomes obvious that for all practical sprays the hydrodynamics at the critical heat flux point are characterized by a closed liquid film and therefore spray impact takes place as drop impact onto a liquid film.

%For the investigation of the influence of the initial temperature on the heat flux density, experiments with different initial temperatures have been performed.

\begin{figure}
	\centering\includegraphics[width=.45\textwidth]{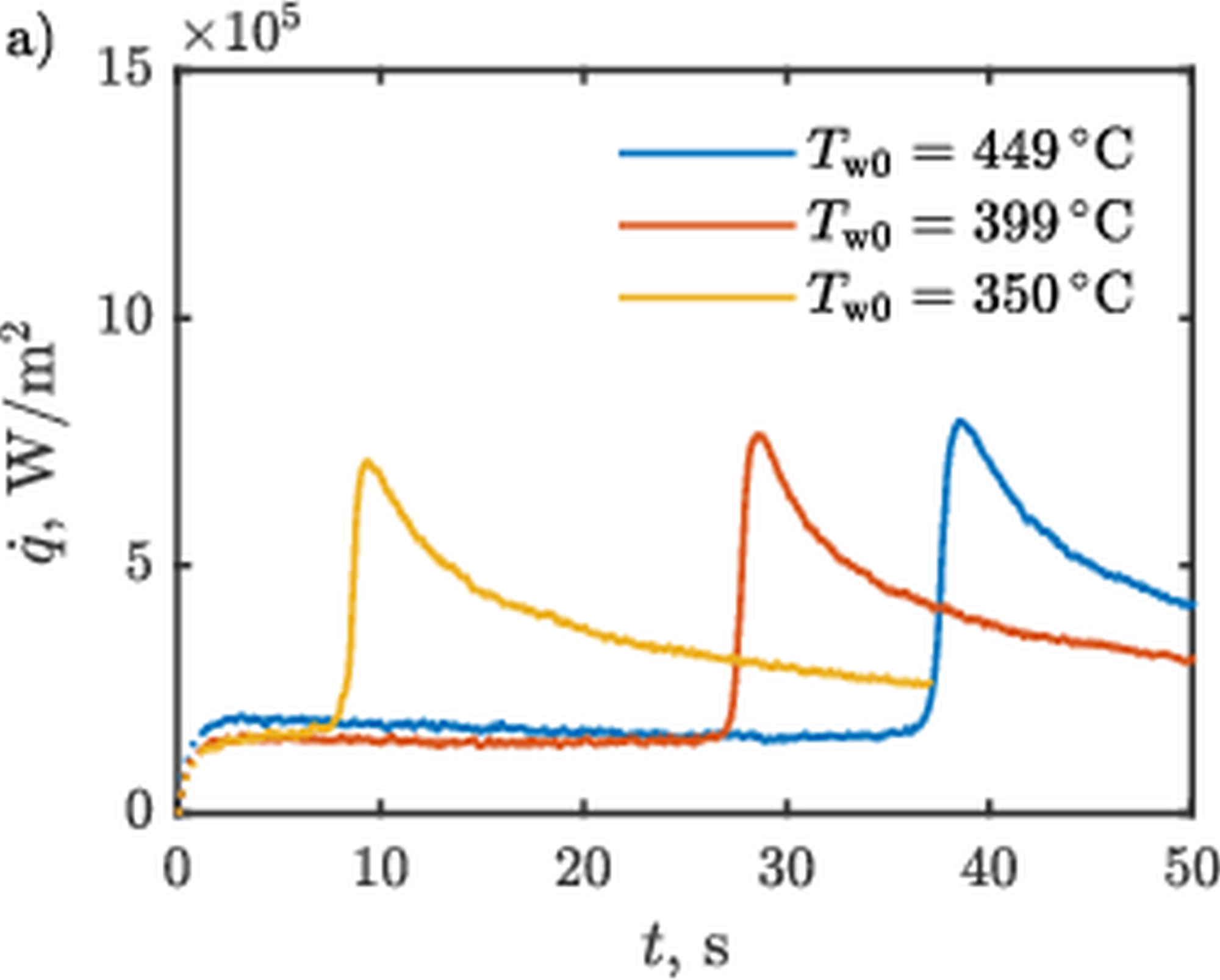}\hspace{0.5 cm}\includegraphics[width=.45\textwidth]{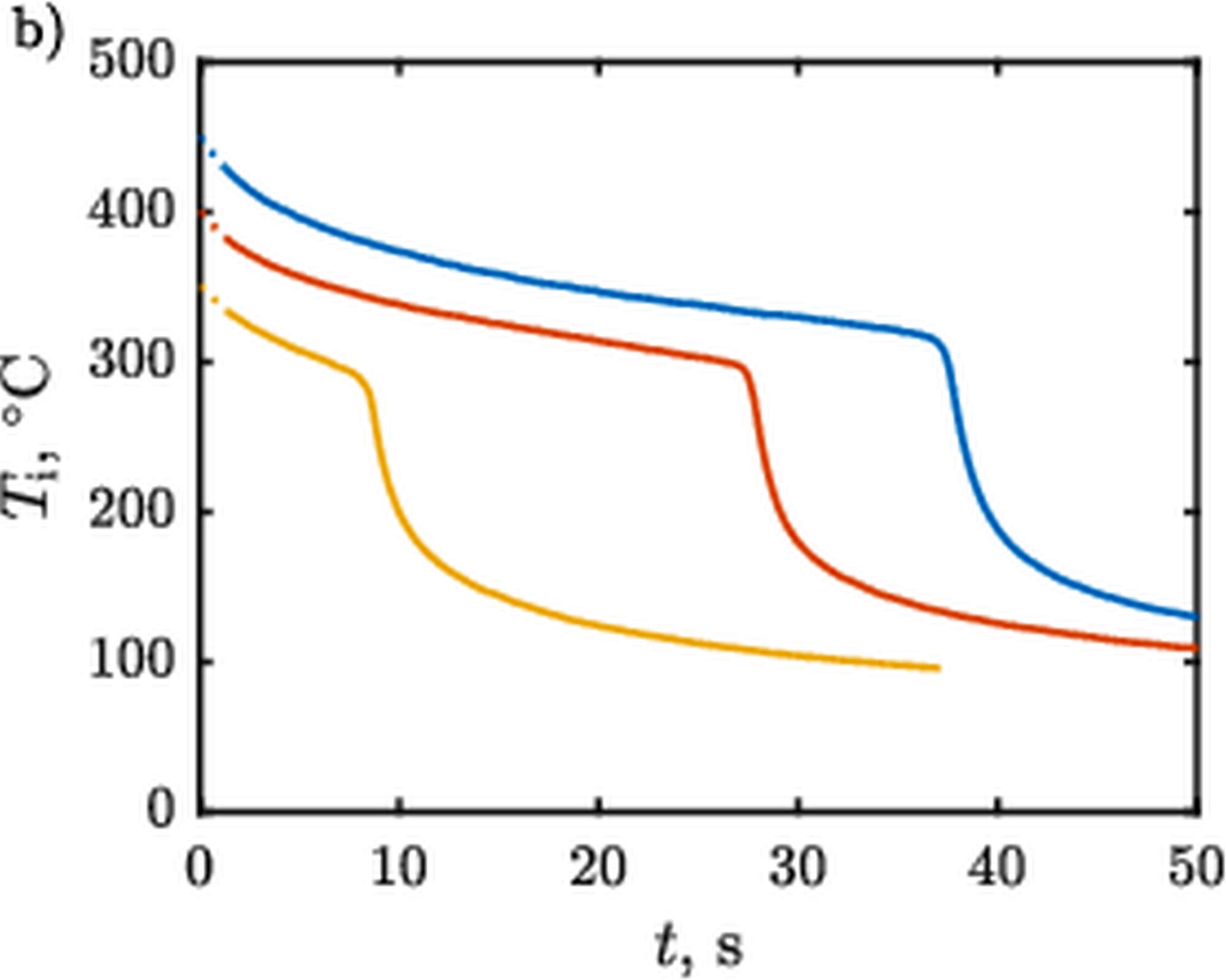}
	\caption{ Influence of the initial wall temperature $T_\mathrm{w0}$ on a) heat flux density and b) surface temperature $T_\mathrm{i}$ dependence on time $t$ for a small mass flux density of $\dot{m}=1.6\, \mathrm{kg/m^2s}$.}
	\label{fig:q_init1}
\end{figure}

 In Fig.~\ref{fig:q_init1} the heat flux density and surface temperature are shown as a function of time for various initial substrate temperatures: $350$, $400$ and $450\, \mathrm{^\circ C}$. The spray properties are: $\dot{m}=1.6\, \mathrm{kg/m^2s}$, $D_{10}=64\, \mathrm{\mu m}$ and $U=8\, \mathrm{m/s}$. To highlight the transient behaviour the temporal axis is limited to $t< 50\,\mathrm{s}$.
Some minor influence of the initial temperature on the heat flux density in the film boiling regime can be identified. However, the overall trend of the curves remain the same. The time shift is a direct result of the different initial substrate temperatures.

\begin{figure}
	\centering\includegraphics[width=.45\textwidth]{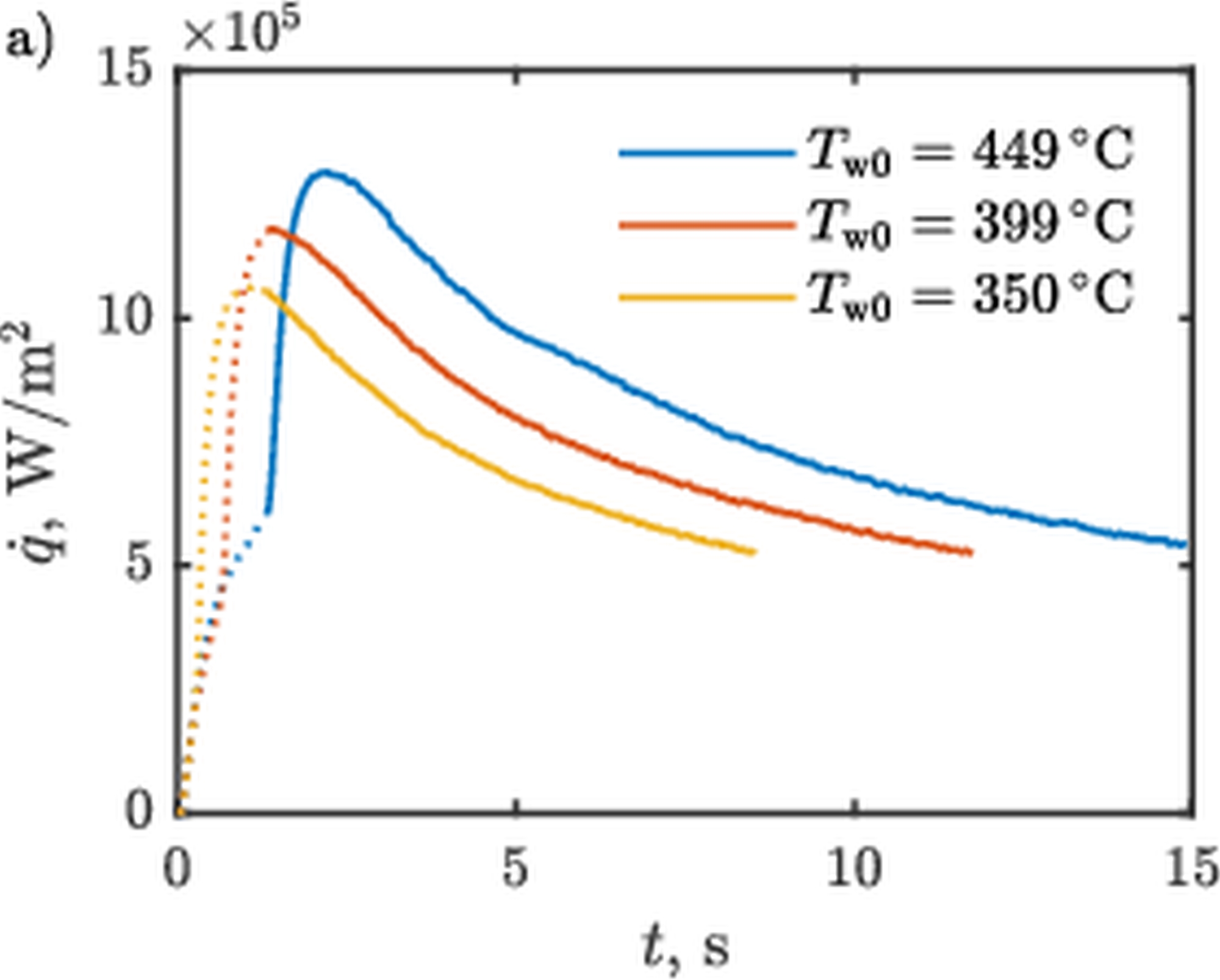}\hspace{0.5 cm}\includegraphics[width=.45\textwidth]{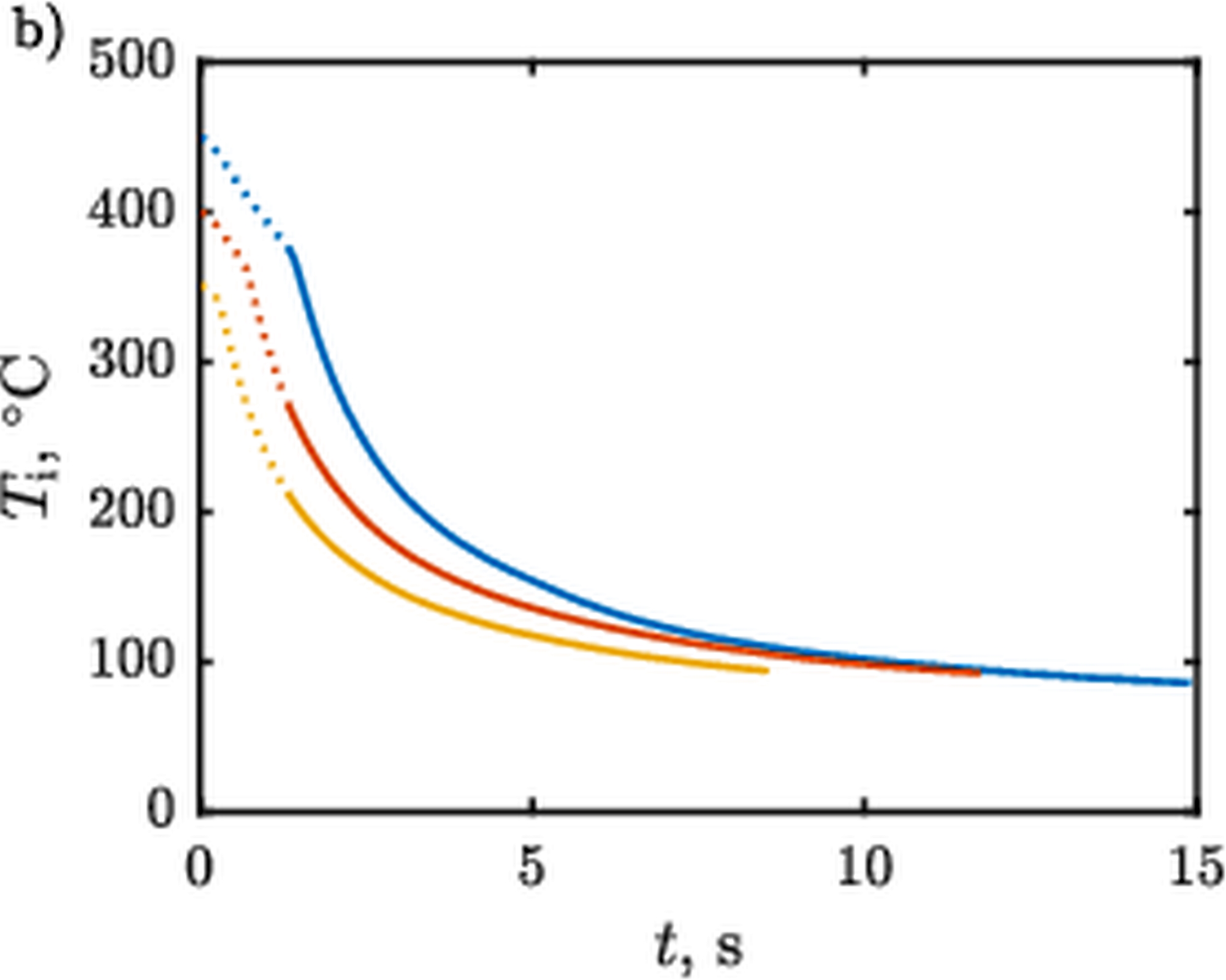}
	\caption{ Influence of the initial wall temperature $T_\mathrm{w0}$ on a) heat flux density and b) surface temperature $T_\mathrm{i}$ dependence on time $t$ for a higher mass flux density of $\dot{m}=9.3\, \mathrm{kg/m^2s}$.}
	\label{fig:q_init2}
\end{figure}

 Similar curves, but this time for a larger mass flux density, are shown in Fig.~\ref{fig:q_init2}. The spray properties are: $\dot{m}=9.3\, \mathrm{kg/m^2s}$, $D_{10}=48\, \mathrm{\mu m}$ and $U=15.6\, \mathrm{m/s}$. Again the slopes are comparable but the heat flux density at the critical heat flux is higher for the higher initial substrate temperatures. Furthermore, no typical film boiling regime is visible. This can again be explained by the small time scales in the film boiling regimes which are of the same order as the rise time of the thermocouples.

%In Fig.~\ref{fig:q_m} b) the heat flux density is shown as a function of the surface temperature for the same mass flux densities  $\dot m = 1.6\, \mathrm{kg/m^2s}$ and $\dot m = 9.3\, \mathrm{kg/m^2s}$ but various initial substrate temperatures: $350$, $400$ and $450\, \mathrm{^\circ C}$. Some minor influence of  the initial temperature on the heat flux density at larger times can be identified. Again there is no minimum of the heat flux density visible in the case of the higher mass flux density due to the limited temporal response of the measurement system and for this reason these portions of the curves are presented as dotted lines.

\begin{figure}
	\centering\includegraphics[width=.45\textwidth]{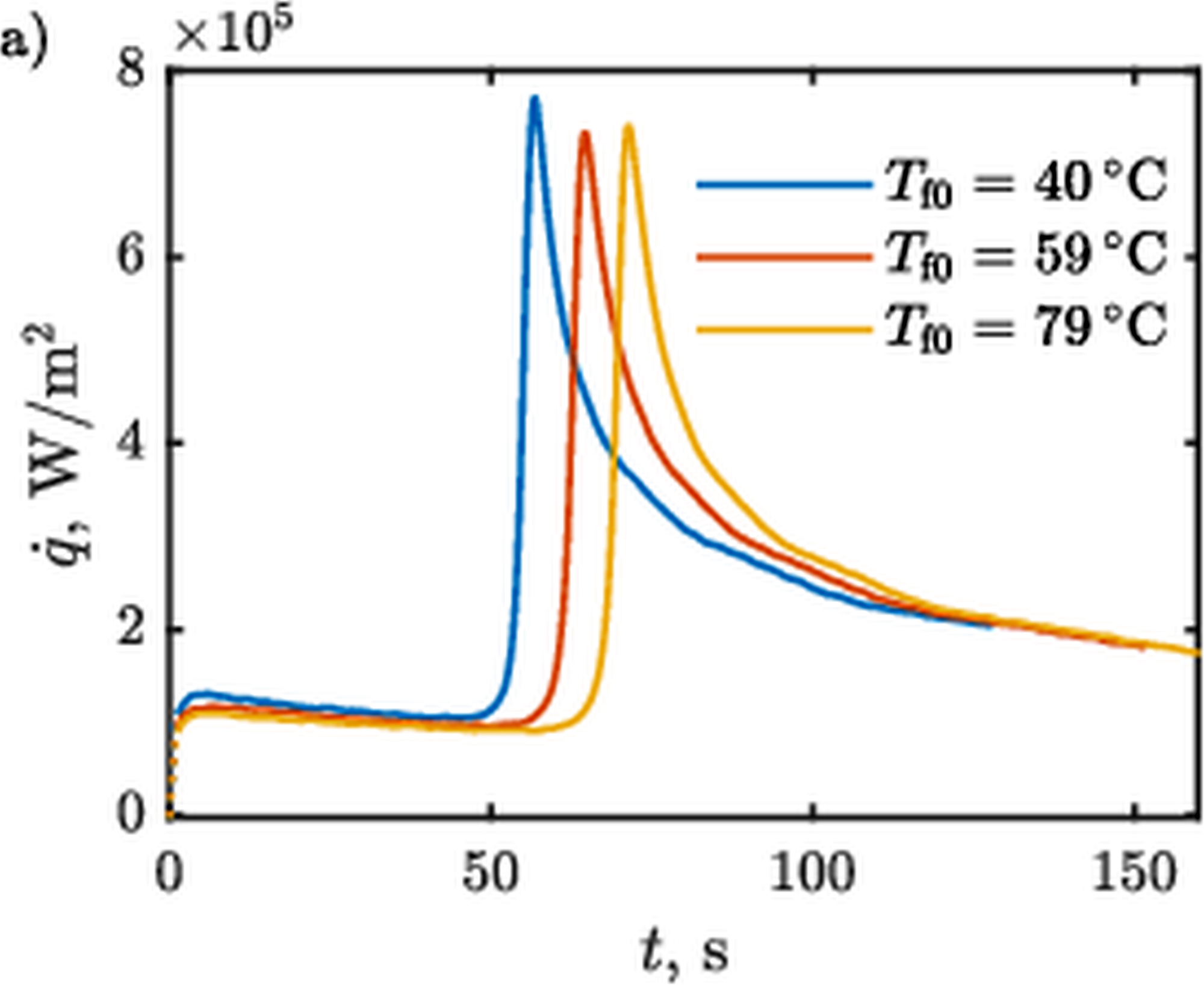}\hspace{0.5 cm}\includegraphics[width=.45\textwidth]{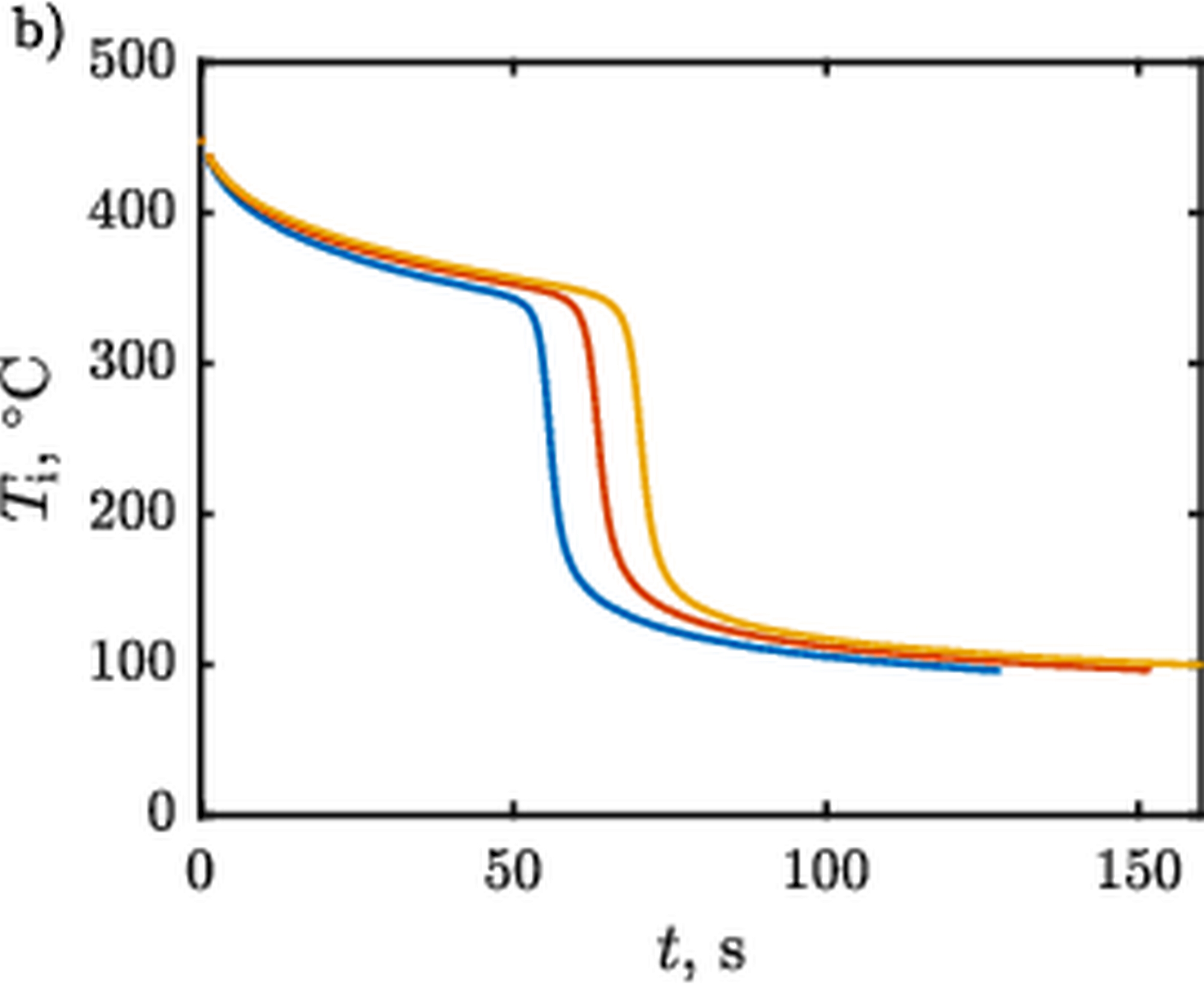}
	\caption{ Influence of the spray fluid temperature $T_\mathrm{f0}$ on a) heat flux density and b) surface temperature $T_\mathrm{i}$ dependence on time $t$.}
	\label{fig:infl_T}
\end{figure}

 Since there is obviously an influence of the spray fluid temperature on the heat flux density, experiments with different spray fluid temperatures $T_\mathrm{f0} = 20-80\, \mathrm{^\circ C}$ were performed. An exemplary result in the known form of heat flux density dependence on the surface temperature is shown in Fig.~\ref{fig:infl_T}. The spray parameters are: $\dot{m}=0.9\, \mathrm{kg/m^2s}$, $D_{10}=43\, \mathrm{\mu m}$ and $U=9.9\, \mathrm{m/s}$. An increasing spray fluid temperature results in a nonlinear decrease of the heat flux density in the film boiling regime. The Leidenfrost point remains nearly constant.

%\begin{figure}
%    \centering
%	\includegraphics[width=.5\textwidth]{q_T_infl_T_t_zoom.jpg}
%	\caption{Heat flux density dependence on the surface temperature for different spray fluid temperatures $T_\mathrm{f0}$.}
%	\label{fig:infl_T_zoom}
%\end{figure}

Heat flux can ``remember'' the initial wall temperature only through the processes occurring in the substrate, since the times associated with drop impacts in the spray are very short. These processes are determined by the heat conduction in a thin thermal boundary layer in the substrate, as analyzed in the next section.

%At temperatures below CHF most data of the same spray parameters but with different initial temperatures converge to one single curve. This means that the effect of the film boiling regime and its duration on the intensive heat transfer in the nucleate boiling regime is rather small.
\section{Analysis of heat transfer during spray cooling}
%\begin{figure}[htbp]
%	\includegraphics[width=\textwidth,keepaspectratio=true]{saturation.png}
%	\caption{Sketch of a hypothetical ideal spray cooling process}
%	\label{fig:saturation}
%\end{figure}

Let us analyze heat transfer during very intensive transient spray cooling of an initially uniformly heated substrate. Consider for simplicity one-dimensional heat conduction in a semi-infinite solid substrate.  This assumption is valid for  cases when the thickness of the thermal boundary layer, $\sqrt{\alpha t}$, is much smaller than the thickness of the target, and the temperature gradients in this boundary are much higher than the gradients associated with the spray distribution in the radial direction. Here $\alpha=\lambda/\rho c_p$ is the thermal diffusivity, where $\lambda$ is the thermal conductivity, $\rho$ is the density and $c_p$ is the heat capacity of the target material.

 In our case the material properties of the substrate are: $\lambda=18\, \mathrm{W/m K}$, $\rho = 7900\, \mathrm{kg/m^3}$ and $c_p = 500\, \mathrm{J/kg K}$. The longest experiments last about $200\, \mathrm{s}$. This results in a thermal boundary layer thickness of about $30\, \mathrm{mm}$ which is comparable to the half of the target height. Therefore the heat conduction in the target can be considered as semi-infinite. Since the spray parameters are nearly constant in the centre area the assumption of a one-dimensional problem is valid.

%The previously shown results, that the heat flux density in the nucleate boiling regime during transient spray cooling becomes independent of the temporal history of the cooling process leads to the development of a simple scaling. As shown in Fig.~\ref{fig:saturation} we start at $t=t_0$ with an homogeneously heated semi-infinite target having the temperature $T_\mathrm{w0}$. At any later time $t>t_0$ the surface is continuously sprayed with an optimal spray. This hypothetically ideal spray has the highest cooling performance that is physically possible. Due to the fact that cooling takes place above saturation temperature, phase change happens during spray impingement. Therefore an ideal spray would result in saturation temperature $T_\mathrm{sat}$ at the surface.

Consider also a coordinate system $\{z,t\}$ fixed at the interface $z=0$ of the semi-infinite target, belonging to the interval $0<z<\infty$. The temperature field $T(z,t)$ in the target can be calculated by solving the one-dimensional energy equation
\begin{equation}
\frac{\partial T}{\partial t} =\alpha \frac{\partial^2 T}{\partial z^2}.
\end{equation}

 Following Duahamel's theorem \cite{Ozsk1980}, the solution which satisfies the boundary condition far from the target interface and the the initial condition
%\begin{eqnarray}
%T &=& T_{w0}, \quad \mathrm{at }\quad t = 0,\quad z \in [0,\infty],\\
%T &=& T_{w0}, \quad \mathrm{at }\quad t>0,\quad z\rightarrow \infty,
%\end{eqnarray}
\begin{equation}
T = T_\mathrm{w0} \quad \mathrm{at }\quad (t = 0\wedge z \in [0,\infty]) \quad \vee \quad  (t>0\wedge  z\rightarrow \infty),
\end{equation}
is
\begin{equation}\label{solgen}
T(t) = T_\mathrm{w0} +\int_0^t A(\tau) \mathrm{erfc}\left[\frac{z}{2\sqrt{\alpha (t-\tau)}}\right]\mathrm{d}\tau,
\end{equation}
where $A(\tau)$ is a function determined by the conditions at the target interface $z=0$, $\mathrm{erfc}$ is the complementary error function and $T_\mathrm{w0}$ is the initial wall temperature. As usual, the limit $z\rightarrow\infty$ denotes  a position at a final distance much larger than the thickness of the thermal boundary layer in the wall. The expression (\ref{solgen}) allows  the general solution for the interface temperature $T_\mathrm{i}(t)$ and for the heat flux density $\dot{q}(t)$ to be determined
\begin{equation}\label{solgenQ1}
T_\mathrm{i}(t) = T_\mathrm{w0} +\int_0^t A(\tau) \mathrm{d}\tau,
\quad
\dot{q}(t) = -\frac{\epsilon_\mathrm{w} }{\sqrt{\pi}} \int_0^t \frac{ T_\mathrm{i}'(\tau)}{\sqrt{t-\tau}} \mathrm{d}\tau,
\end{equation}
where $\epsilon_\mathrm{w} = \sqrt{\lambda \rho c_p}$ is the thermal effusivity of the wall.
%This solution yields the following relation between the interface temperature and the heat flux density:
%\begin{equation}
%    \dot{q}(t) = -\frac{\epsilon_w }{\sqrt{\pi}} \int_0^t %\frac{T_\mathrm{i}'(\tau)}{\sqrt{t-\tau}} \mathrm{d}\tau. \label{solgenQ}
%\end{equation}

%In a particular case when the surface temperature jumps at $t=0$ immediately to the saturation temperature $T_\mathrm{sat}$ the function $A$ in (\ref{solgenT}) is $A=T_\mathrm{sat}-T_{w0}\delta(\tau)$, where $\delta(\tau)$ is the Dirac delta function, leads to the following expression for the heat flux density:
%\begin{equation}\label{eq:limit0}
%\dot{q}(t)=\frac{\epsilon_w}{\sqrt{\pi}}  \frac{\Delta T}{\sqrt{t}},\quad \Delta T=T_\mathrm{w0}-T_\mathrm{sat}.
%\end{equation}

\subsection{Temperature and heat flux evolution in the film boiling regime}

%The model  for the heat flux in the film boiling regime agrees very well with the experimental data.

 The theoretical model \cite{Breitenbach2017} for the heat transfer from a single drop and during spray cooling in the film boiling regime
is based on the analysis of the heat conduction in the substrate, heat convection in the liquid region and in an expanding thin vapor layer emerging between the impacting drop and the very hot substrate. The mass flux of the vapor generated at the lower liquid interface is determined from the energy balance at this interface.
This model has already been validated by comparison with experimental results from  literature \cite{Wendelstorf2008}. Moreover, the predicted vapor layer thickness agrees well with the direct measurements of \cite{Tran2012} and \cite{Chaze2019}. Predictions for the evolution of the heat flux density $\dot q(t)$ also agree with the accurate measurements based on the infrared technique \cite{Chaze2019}.

 The total heat transferred during the impact of a single drop  $Q_\mathrm{single}$ is determined by the integration of the heat flux density $\dot q(t)$ over the ``apparent contact area'' during the contact time. It should be noted that the contact area cannot be based on the drop spreading diameter since the free lamella in the remote regions can levitate \cite{Roisman2018}.  Therefore the values of $D_0^2$ and $D_0/U_0$ are used as the scales for the contact area and for the contact duration, where $U_0$ and $D_0$ are the impact velocity and drop diameter, respectively.

This analysis \cite{Breitenbach2017} allows  the heat flux density during spray impact in the film boiling regime to be predicted
\begin{eqnarray}
    \dot{q}&=&S \epsilon_\mathrm{w}  (T_\mathrm{i}-T_\mathrm{sat}),\label{qdt}\\
    S &=& 8.85\chi \frac{\dot{m}}{\rho_\mathrm{f} D_{10}^{1/2} U^{1/2}\left[1-b+\sqrt{(1-b)^2+w}\right]},\label{qFilmBoil}\\
    w &=& \frac{8 (T_\mathrm{i}-T_\mathrm{sat})\epsilon_\mathrm{w}^2}{\pi \lambda_\mathrm{v} \rho_\mathrm{f} L}, \quad b = \frac{2\sqrt{5} \epsilon_\mathrm{w} \epsilon_\mathrm{f} (T_\mathrm{sat}-T_\mathrm{f0})}{\pi \rho_\mathrm{f} \lambda_\mathrm{v} L},\label{W_b}
\end{eqnarray}
where $L$ is the latent heat of evaporation, $\chi$ is a dimensionless fitting parameter which depends on the substrate wetting properties and roughness, $T_\mathrm{sat}$ is the saturation temperature of the liquid and $T_\mathrm{f0}$ is the initial spray fluid temperature. All the terms with the subscript ``f'' correspond to the liquid (fluid) component, ``w'' to the wall and ``v'' to the vapor.

 Since the sprays used in the experiments and  in  practical applications are always polydisperse, the average drop diameter and velocity, $D_{10}$ and $U$, are used in the model. The parameter $\chi$ thus inherently accounts also for the drop size and velocity distributions.

 In this study the model (\ref{qdt})-(\ref{W_b}) is used for prediction of the evolution of the wall temperature in time. The predictions are then compared with our experimental data.

%In our experimental conditions the estimated values of $w$ and $b^2$ are both of order $10^2$. Since the temperatures associated with the film boiling regime are well above the saturation temperature, the value $(T_\mathrm{i}-T_\mathrm{sat})/(T_\mathrm{w0}-T_\mathrm{sat})\approx 1$. The value of $S$  can be estimated by using $T_\mathrm{w0}$ instead of $T_\mathrm{i}(t)$ in the expression for $w$ in (\ref{W_b}).

%Besser diesen Teil verweden?
 In our experimental conditions, $T_\mathrm{i} = 340\, \mathrm{^\circ C}$, $T_\mathrm{sat} = 99\, \mathrm{^\circ C}$, $\epsilon_\mathrm{w} = 8432\, \mathrm{J/Km^2s^{1/2}}$, $\lambda_\mathrm{v} = 0.0248\, \mathrm{W/mK}$, $\rho_\mathrm{f} = 998\, \mathrm{kg/m^3}$, $L = 2453\, \mathrm{kJ/kg}$, $\epsilon_\mathrm{f}  = 1581\, \mathrm{J{/}Km^2s^{1/2}}$, $T_\mathrm{f0} = 20\, \mathrm{^\circ C}$, the estimated values for $w$ and $b$ defined in (\ref{W_b}) are $w \approx 700$ and $b\approx 25$. Since $b\gg 1$ and  $b^2$ and $w$ are of the same order of magnitude, the effect of the dependence of $S$ on the changing temperature $T_\mathrm{i}$ in the expression for $w$ can be neglected. The value of $S$ can be estimated by using $T_\mathrm{w0}$ instead of $T_\mathrm{i}(t)$ in the expression (\ref{W_b}).

Expressions (\ref{solgenQ1}) and (\ref{qdt}) lead to the following integral equation for the surface temperature, presented in the dimensionless form
%\begin{equation}
%    S \epsilon_\mathrm{w} (T_\mathrm{i}(t)-T_\mathrm{sat}) + \frac{\epsilon_w }{\sqrt{\pi}} \int_0^t \frac{T_\mathrm{i}'(\tau)}{\sqrt{t-\tau}} \mathrm{d}\tau=0.
%\end{equation}
%This equation can be rewritten in  dimensionless form
\begin{eqnarray}
    \Theta(\xi) &+&  \int_0^\xi \frac{\Theta'(\zeta)}{\sqrt{\xi-\zeta}} \mathrm{d}\zeta=0,\label{solutionTheta}
    \end{eqnarray}
    where the surface temperature is made dimensionless using
    \begin{eqnarray}
    \Theta &=& \frac{T_\mathrm{i}(t)-T_\mathrm{sat}}{T_\mathrm{w0}-T_\mathrm{sat}}, \quad \xi = t \pi S^2, \quad \zeta = \tau \pi S^2. \label{scaling}
\end{eqnarray}
This equation can be solved numerically subject the initial condition $\Theta(0)=1$. The analytical solution for $\Theta(\xi)$ can be represented as a series
\begin{eqnarray}
    \Theta(\xi) &=& 1 + \sum_{i=1}^\infty a_i \xi^{i/2}, \label{ThetaTheor}\\
    a_1 &=&  -\frac{2}{\pi},\quad a_2=\frac{1}{\pi},\quad a_3 = -\frac{4}{3\pi^2}, \quad ...\quad
    a_{i+1} = -a_i \frac{2^{-i} \Gamma(i+1)}{\Gamma\left[\frac{i+1}{2}\right]\Gamma\left[\frac{i+3}{2}\right]}
\end{eqnarray}
where $\Gamma$ is the gamma function.

The corresponding heat flux density $\dot{q}$ can be estimated using  (\ref{solgenQ1})
\begin{eqnarray}
    \dot{q}&=& S \epsilon_\mathrm{w} (T_\mathrm{w0}-T_\mathrm{sat}) \Phi(\xi),\label{qdottheta}\\
    \Phi(\xi)&=& - \int_0^\xi  \frac{\Theta'(\zeta)}{\sqrt{\xi-\zeta}}\mathrm{d}\zeta,\quad \Theta'(\zeta)=\frac{1}{2}\sum_{i=1}^\infty a_i i \zeta^{i/2-1}.\label{PhiTheor}
    \end{eqnarray}

%The numerical solution for the dimensionless surface temperature $\Theta(\xi)$, its derivative $\Theta'(\xi)$ and the dimensionless heat flux density $\Phi(\xi)$ is shown in Fig.~\ref{fig:ThetaPhiNum}

%\begin{figure}
%	\center\includegraphics[width=0.49\textwidth,keepaspectratio=true]{ThetaTheory.jpg}
%	\includegraphics[width=0.49\textwidth,keepaspectratio=true]{PhiTheory.jpg}
%	\caption{Theoretically predicted evolution of the dimensionless substrate temperature $\Theta(\xi)$, defined in (\ref{scaling}), its derivative $\Theta'(\xi)$ and the dimensionless heat flux $\Phi(\xi)$ as functions of the dimensionless time $\xi$ in the film boiling regime. The solution, based on the  first 50 terms in the series (\ref{ThetaTheor}), converges on the interval $0<\xi<22$.
%	\label{fig:ThetaPhiNum}
%\end{figure}

%In Fig.~\ref{fig:ThetaTheorExp} the measured evolution of the dimensionless surface temperature $\Theta(\xi)$ is plotted as a function of  dimensionless time $\xi$ and is compared with the theoretical solution  (\ref{ThetaTheor}). The computation of the theoretical solution is  based on the  first 50 terms in the series. It converges on the interval $0<\xi<22$.

%For the reduction of the experimental data, $\chi$ in (\ref{qFilmBoil}) is fitted to the experimental data using a least square fit, resulting in $\chi=2$. The experimental parameters corresponding to the shown experimental data span  the following ranges: $\dot{m}=0.5 - 3.7\, \mathrm{kg/m^2s}$, $D_{10}=43 - 78\, \mathrm{\mu m}$, $U=6.7 - 12.8\, \mathrm{m/s}$ and $T_\mathrm{w0}=350 - 450 \, \mathrm{^\circ C}$.

\begin{figure}
    \centering
	\includegraphics[width=.5\textwidth]{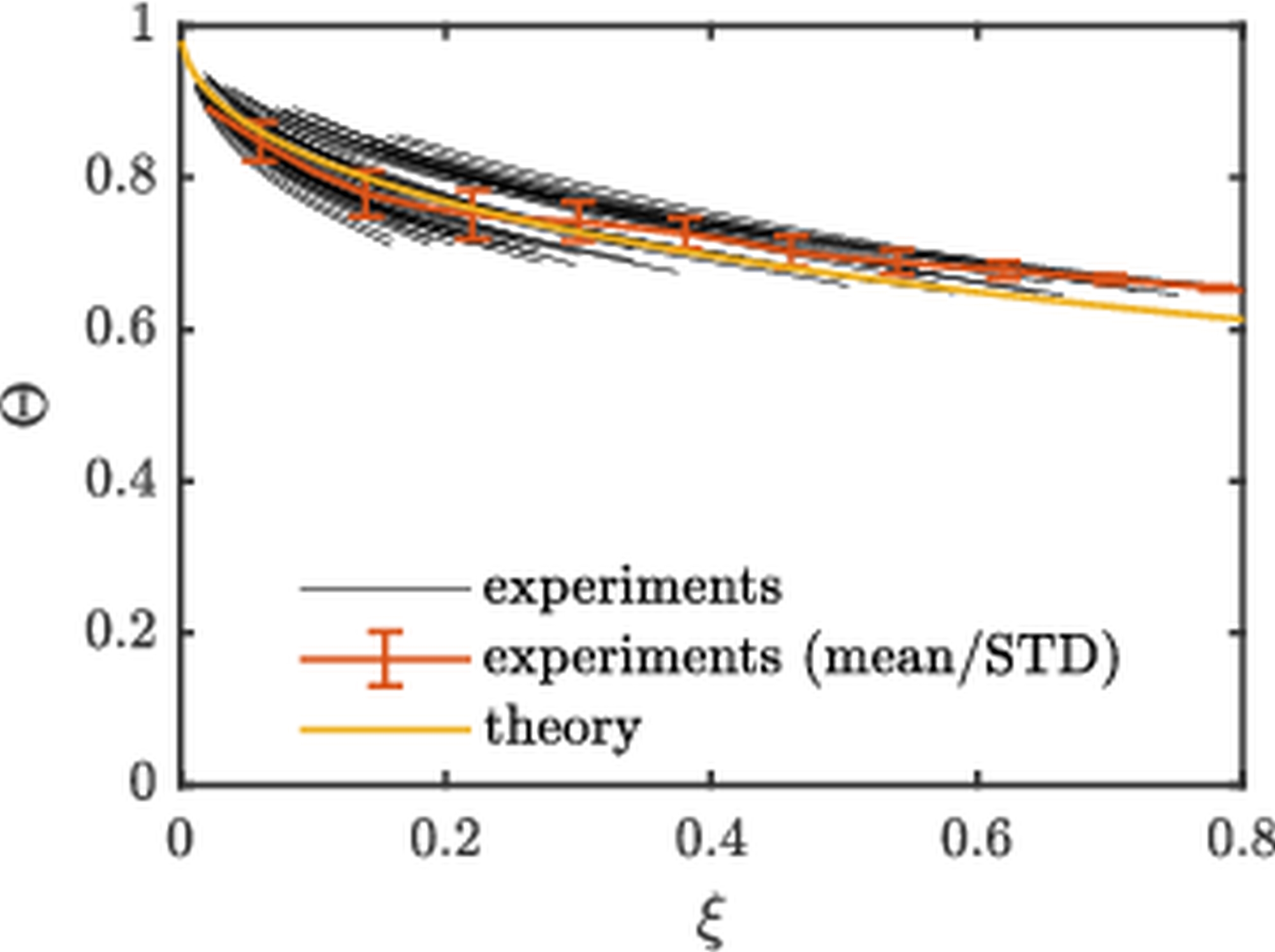}
	\caption{Evolution of the dimensionless substrate temperature $\Theta(\xi)$, defined in (\ref{scaling}) as a function of the dimensionless time $\xi$ for the experimental data  in the film boiling regime compared with the theoretical prediction obtained by integration of  (\ref{solutionTheta}). }
	\label{fig:ThetaTheorExp}
\end{figure}

In Fig.~\ref{fig:ThetaTheorExp} the measured evolution of the dimensionless surface temperature $\Theta(\xi)$ as a function of dimensionless time $\xi$ is plotted for the theoretical prediction (\ref{ThetaTheor}) and for experimental results. The computation of the theoretical solution is  based on the  first 50 terms in the series. It converges on the interval $0<\xi<22$.  The line shown for the experiments is the mean and the error bars indicate the standard deviation computed over $49$ experiments. For the reduction of the experimental data $T_\mathrm{i}$ and $t$, $\chi$ in (\ref{qFilmBoil}) is fitted to the experimental data using a least square fit, resulting in $\chi=2.2$. Only the experiments exhibiting clear film boiling behaviour are chosen and  the experimental data comprising the film boiling regime are plotted. We skip the experiments showing no film boiling behaviour because of the previously mentioned limited temporal response of the measurement system. The experimental parameters corresponding to the shown experimental data span the following ranges: $\dot{m}=0.5 - 9.1\, \mathrm{kg/m^2s}$, $D_{10}=43 - 78\, \mathrm{\mu m}$, $U=6.7 - 15.9\, \mathrm{m/s}$, $T_\mathrm{w0}=350 - 450 \, \mathrm{^\circ C}$  and $T_\mathrm{f0}=18-80\, \mathrm{^\circ C}$. Although the parameter range in this study is quite small, the authors in \cite{Breitenbach2017} show that the model is valid for a much lager parameter span.

\begin{figure}
    \centering
	\includegraphics[width=.5\textwidth]{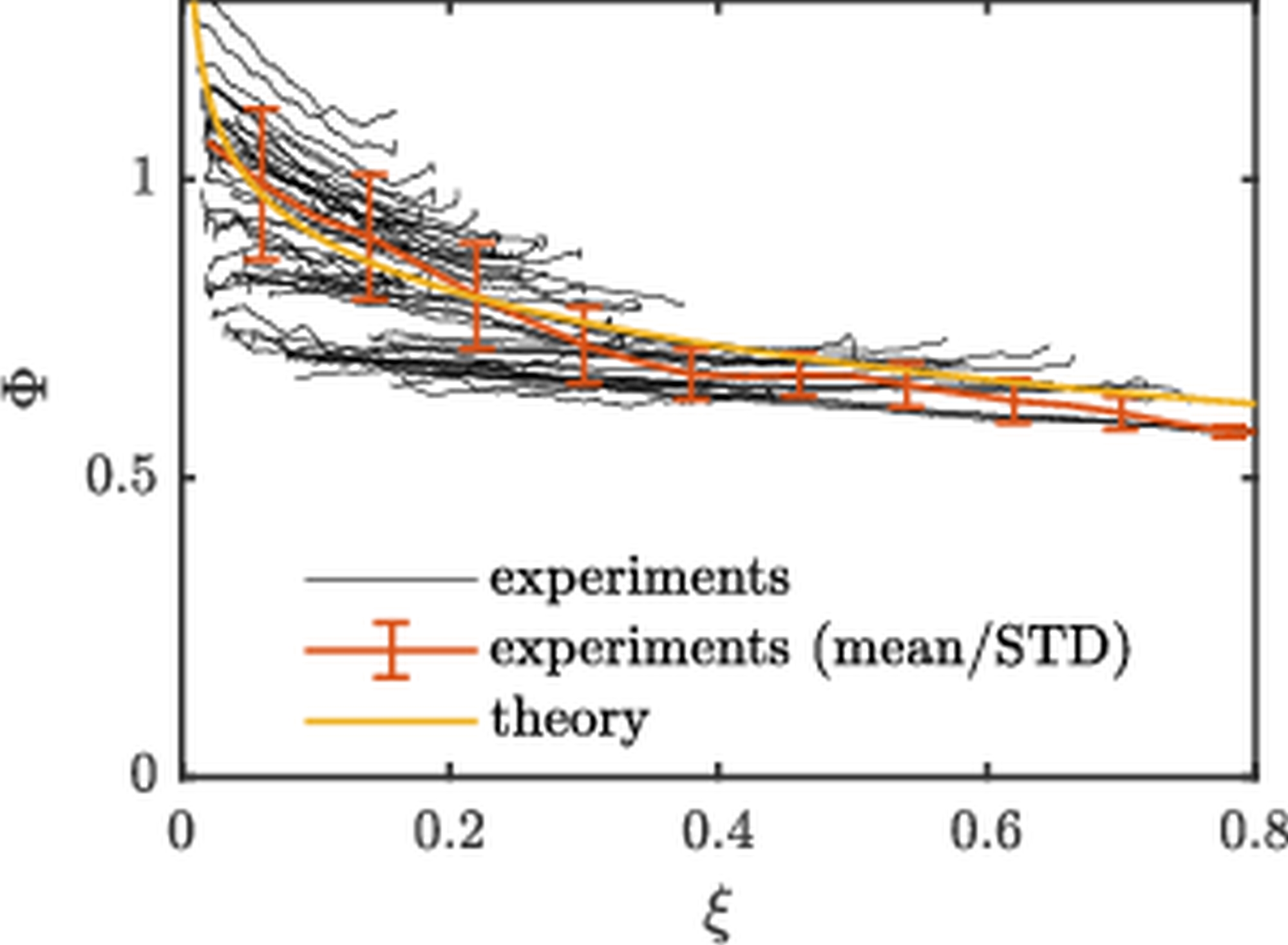}
	\caption{ Evolution of the dimensionless substrate heat flux density $\Phi(\xi)$, defined in (\ref{qdottheta}) as a function of the dimensionless time $\xi$ for experimental data  in the film boiling regime and  for the theoretical solution (\ref{PhiTheor}).}
	\label{fig:PhiTheorExp}
\end{figure}

 In Fig.~\ref{fig:PhiTheorExp}  the corresponding evolution of the dimensionless heat flux density for the same experiments is shown and compared  to  theory. Although there is obviously a large scatter in the experimental data,  very good agreement between the experiments and  theory can be observed. Especially $\chi$, being close to unity, indicates the performance of the theory, which obviously captures all main physical players. The shown data consists of different sources like phase Doppler, patternator and heat flux data, each having numerous sources of uncertainty. Keeping that in mind, the scatter is  acceptable and therefore the good agreement between experiment and theory indicates a good understanding of the physics of spray cooling in the film boiling regime.


\subsection{ Transitional spray cooling regime below the Leidenfrost point}
 At some instant $t_\mathrm{L}$ the conditions at the surface correspond to the Leidenfrost point. We identify the Leidenfrost point from the time series of the measurement data as the point when the heat flux density reaches its minimum in the film boiling regime. The value of the surface temperature at the Leidenfrost point in our study is not a fixed value but variable. This temperature depends not only on the impact parameters of the drops in the spray or mass flux density, but probably on the rate of the wall cooling in the film boiling regime. At this stage it is not easy to develop a complete and reliable model for the Leidenfrost temperature. In our experiments it is in the range $315-360\, \mathrm{^\circ C}$. At this stage we leave the exact modeling of the Leidenfrost temperature for future studies and focus on the description of the nucleate boiling at lower temperatures.

 Denote the surface temperature at the Leidenfrost point as $T_\mathrm{iL}$. The corresponding heat flux density $\dot{q}_\mathrm{L}$ can be estimated using (\ref{qdt}).

Further cooling leads to  a rapid increase of the heat flux density, caused by the partial wetting of the surface. The area of the wetted spots quickly grows. The duration of this transitional spray cooling regime is rather short. At this instant both modes of drop impact can be observed: drop impact onto dry regions in the film boiling regime and drop impact onto wetted spots. The relative area of the wetted spots  grows with increasing time; hence, with decreasing surface temperature.

\begin{figure}
    \centering
	\includegraphics[width=.5\textwidth]{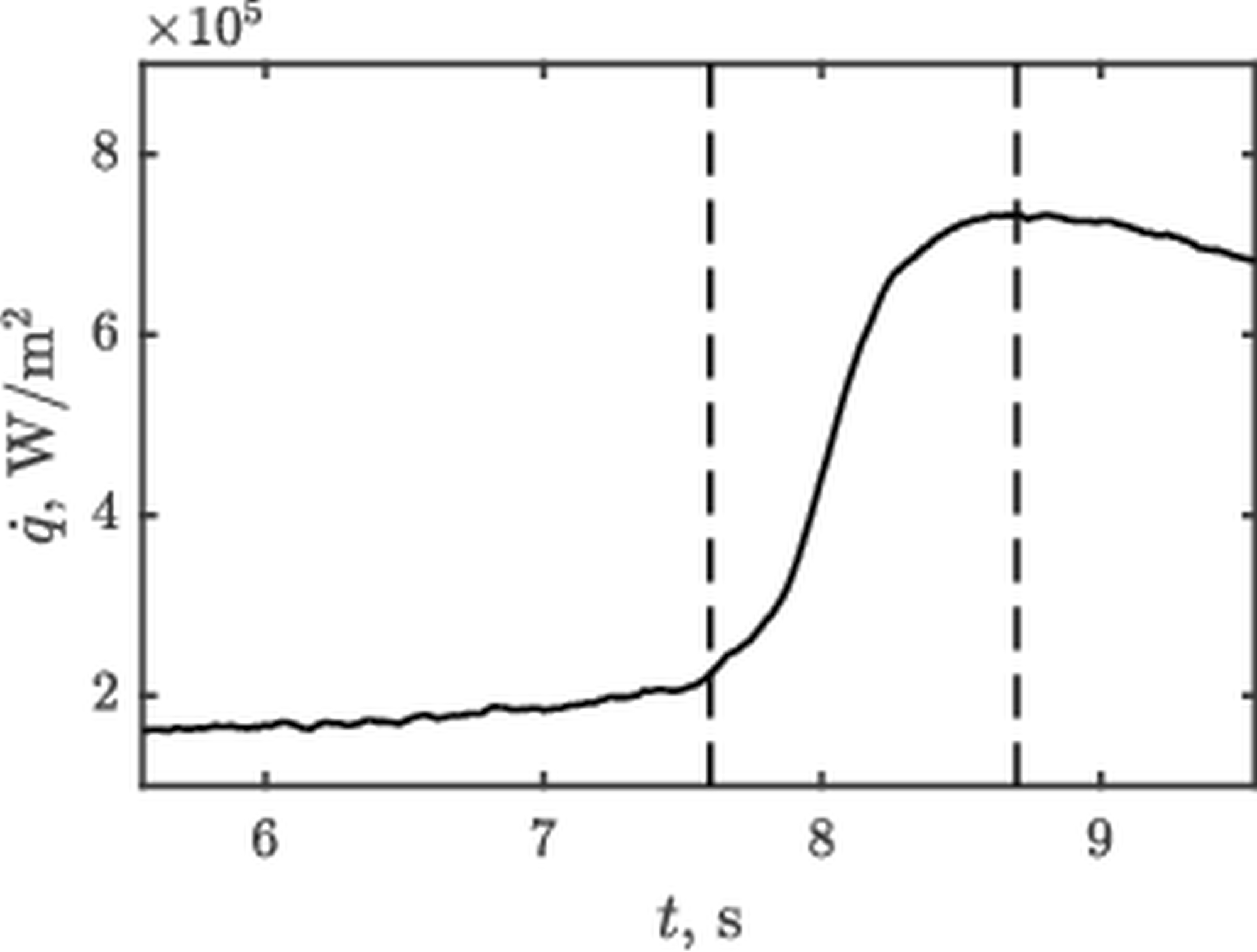}
	\caption{ Evolution of the heat flux density as a function of time for an exemplary experiment. The parameters of the spray are: $\dot{m}=1.4\, \mathrm{kg/m^2s}$, $D_{10}=43\, \mathrm{\mu m}$ and $U=11\, \mathrm{m/s}$. The dashed vertical lines bound the transition boiling regime.}
	\label{fig:qtExemplTrans}
\end{figure}

 In Fig.~\ref{fig:qtExemplTrans} an exemplary dependence of the heat flux density  on time is shown. Here we limit the temporal span to the time between the end of the film boiling regime and the beginning of the nucleate boiling regime in order to illustrate the behaviour in the transition boiling regime. The corresponding spray parameters are: $\dot{m}=1.4\, \mathrm{kg/m^2s}$, $D_{10}=43\, \mathrm{\mu m}$ and $U=11\, \mathrm{m/s}$. We can clearly identify the instant when the Leidenfrost point occurs  ($t_\mathrm{L}$) at around $7.6\, \mathrm{s}$ and  the instant when the critical heat flux is reached ($t_\mathrm{CHF}$) at $8.7\, \mathrm{s}$. This small time of slightly more than $1\, \mathrm{s}$ between both events is similar to the rise time of the thermocouples. We obtain no reasonable temperature readings in the range between both events and therefore can make no prediction of the surface temperature and the heat flux density in the transition boiling regime or the critical heat flux.

 This short time in the transition boiling regime is very small compared to the overall cooling process and therefore the heat which is transferred during this short time can be neglected. In the present case the transition boiling regime therefore plays no role with respect to the cooling process. Having this short time in mind we can image the transition boiling regime to be non-existent and replaced by a jump of the heat flux density towards infinity at the Leidenfrost point. The typical trend we see in our measurements of heat flux density is only due to the limited temporal resolution of the measurement equipment and is probably only valid for very low mass flux density or cooling rates in general.

\subsection{Nucleate boiling regime, a remote asymptotic solution}
% At some further instant $t_\mathrm{M}$ the heat flux reaches maximum. At times $t>t_\mathrm{M}$ the cooling occurs in the fully developed nucleated boiling regime. In this regime the heat transfer is influenced by the local superheating of the liquid film which at some instant follows by the periodic creation of the bubbles. The local temperature of the substrate surface oscillates with time. Two typical temperatures characterize the evolution of the substrate temperature. One is the saturation temperature $T_\mathrm{sat}$ in the neighborhood of the multiples contact lines of the expanding bubbles. The second typical temperature is the contact temperature during the waiting time for bubble nucleation (after departure of the previous bubble)
%\begin{equation}
%    T_\mathrm{cont}=\frac{\epsilon_f T_f+\epsilon_w T_{i}}{\epsilon_f %+\epsilon_w}
%\end{equation}

%The thickness of the thermal boundary layer, associated with the oscillations of the typical frequency $\omega$, in this case can estimated as $h_\mathrm{thermal}\sim \sqrt{\alpha/\omega}$. The associated heat flux is therefore $\dot{q}_\omega\sim \lambda (T_\mathrm{w0}-T_\mathrm{i})/h_\mathrm{thermal}\sim \epsilon_\mathrm{w} (T_\mathrm{w0}-T_\mathrm{i}) \sqrt{\omega}$.

 For longer times $t>t_\mathrm{L}$, during the nucleate boiling regime of spray cooling, the heat flux density can be estimated from (\ref{solgenQ1}) accounting for the very short duration of the transient boiling regime and very high temperature time derivative during this regime. Denote $\Delta T_\mathrm{L}$ as the temperature jump during the transition boiling. In this case the expression (\ref{solgenQ1}) yields
\begin{equation}\label{qNucleategov}
    \dot{q}(t) = -\frac{\epsilon_\mathrm{w} }{\sqrt{\pi}} \int_0^{t_\mathrm{L}} \frac{ T_\mathrm{i,film}'(\tau)}{\sqrt{t-\tau}} \mathrm{d}\tau
    + \frac{\epsilon_\mathrm{w}}{\sqrt{\pi}}
    \frac{\Delta T_\mathrm{L}}{\sqrt{t-t_\mathrm{L}}}
    -\frac{\epsilon_\mathrm{w} }{\sqrt{\pi}} \int_{t_\mathrm{L}}^t \frac{ T_\mathrm{i,nucleate}'(\tau)}{\sqrt{t-\tau}} \mathrm{d}\tau .
\end{equation}

 On the right-hand side of (\ref{qNucleategov}) the first term is associated with the thermal history during the film boiling regime, the second term is associated with the temperature jump $\Delta T_\mathrm{L}$ during the transition regime at $\tau=t_\mathrm{L}$, and the last term is based on the temperature evolution during the nucleate boiling at times $t_\mathrm{L}<\tau < t$.

 In order to model the heat flux density in the nucleate boiling regime the values of $\Delta T_\mathrm{L}$ and the evolution of the surface temperature $T_\mathrm{i}(\tau)$ (which is required for the computation of the time derivative $T'_\mathrm{i}(\tau)$) are necessary.

 In the estimation of an upper bound for heat flux density during the nucleate boiling regime of single drop impact \cite{breitenbach2017a} the temperature at the wetted part of the wall interface is approximated by the saturation temperature $T_\mathrm{sat}$.

 The nucleate boiling regime is characterized by an intensive nucleation and expansion of vapor bubbles. The temperature in the vicinity of the contact line of the each expanding bubble is close to the saturation temperature. However, some overheating of the surrounding liquid is required for the bubble growth. The heat transfer in the liquid phase during nucleate boiling regime is governed by the convection in the liquid flow between the bubbles. The heat mainly goes into vaporization at the bubble interfaces where $T=T_\mathrm{sat}$. Therefore, the upper bound for the heat flux during nucleate boiling regime can be estimated by the assumption that the temperature at the wetted wall interface is $T_\mathrm{sat}$. Surprisingly the duration of a single drop evaporation in the nucleate boiling regime, predicted using this rough estimation of $\dot q$, agrees very well with  numerous experimental data \cite{Abu-Zaid2004,Itaru1978,Tartarini1999}.

 In this study the upper bound for the flux density $\dot q$ during spray cooling is  estimated also, as in the case of a single drop impact, using the assumption that the substrate temperature is equal to the saturation temperature at $t>t_\mathrm{L}$. The third term in the right-hand-side of the equation (\ref{qNucleategov}), associated with the time gradient of the surface temperature at $t>t_\mathrm{L}$, can be neglected in comparison to the effect of the temperature jump at the Leidenfrost point. The temperature jump during the transition boiling regime can be estimated as  $\Delta T_\mathrm{L}=T_\mathrm{iL}-T_\mathrm{sat}$.

 The heat flux density can be obtained from (\ref{qNucleategov}), neglecting the value of the last term, in the form
\begin{equation}
    \dot{q} = \frac{\epsilon_\mathrm{w}}{\sqrt{\pi}}
    \frac{T_\mathrm{iL}-T_\mathrm{sat}}{\sqrt{t-t_\mathrm{L}}}
    -S \epsilon_\mathrm{w}\frac{T_\mathrm{w0}-T_\mathrm{sat}}{2} \sum_{i=1}^\infty a_i i  \mathrm{B} \left[\frac{\xi_\mathrm{L}}{\xi};\frac{i}{2},\frac{1}{2}\right]\xi^{\frac{i-1}{2}},\label{qnucleate}
\end{equation}
where $\mathrm{B}[\cdot; \cdot,\cdot]$ is the incomplete beta function. Expression (\ref{qnucleate}) is valid only for very fast substrate cooling, when the time interval between the Leidenfrost point and the point corresponding to the critical heat flux is very short.  It can be further modified using (\ref{scaling})
\begin{equation}
    \dot{q} =\frac{\epsilon_\mathrm{w}}{\sqrt{\pi}}\frac{T_\mathrm{w0}-T_\mathrm{sat}}{\sqrt{t-t_\mathrm{L}}}\left[\Theta(\xi_L)
    -\frac{S\sqrt{\pi}\sqrt{t-t_\mathrm{L}}}{2} \sum_{i=1}^\infty a_i i  \mathrm{B} \left[\frac{\xi_\mathrm{L}}{\xi};\frac{i}{2},\frac{1}{2}\right]\xi^{\frac{i-1}{2}}\right],\label{Qnucleate}
\end{equation}

Moreover, at large times, $t\gg t_\mathrm{L}$, the expression (\ref{Qnucleate}) approaches the following remote asymptotic solution
\begin{equation}
    \dot{q} \approx \frac{\epsilon_\mathrm{w}}{\sqrt{\pi}}\frac{T_\mathrm{w0}-T_\mathrm{sat}}{\sqrt{t-t_\mathrm{L}}}\left[\Theta(\xi_L)
    +\frac{2 S \sqrt{t_\mathrm{L}}}{\sqrt{\pi}}\right].\label{QnucleateRem}
\end{equation}

%For long times and moderate initial wall temperatures only two first dominant terms in the series (\ref{qnucleate}) can be taken. The approximation of for the heat flux yields
%\begin{equation}
%    \dot{q} = \chi \frac{\epsilon_\mathrm{w}}{\sqrt{\pi}}
%    \frac{T_\mathrm{iL}-T_\mathrm{sat}}{\sqrt{t-t_\mathrm{iL}}}
%    -S \frac{T_\mathrm{w0}-T_\mathrm{sat}}{2}
%\end{equation}

%the present case the boundary conditions are of Dirichlet type: $T(z=0)=T_\mathrm{sat}$ at the top and $T(z=\infty)=T_\mathrm{w0}$ at infinity.
%This problem can be soled by means of similarity analysis \cite{ROISMAN2010} resulting in for the temperature with the similarity variable $\xi = \frac{z}{\sqrt{\alpha t}}$ and $\Delta T =T_\mathrm{w0}-T_\mathrm{sat}$. The heat flux density at the top surface is calculated from the derivation of the surface temperature with respect to $z$ multiplied by the thermal conductivity $k$:

% As long as the surface temperature is high enough that means phase change during spray cooling is the dominant mechanism, there is no physical possibility to reach a lower temperature than $T_{sat}$ at the surface. Therefore equation~\ref{eq:limit} can be considered as a upper limit of the heat flux density which can be achieved with spray cooling. This limit is only dependent on the material's thermal properties $e_w$, time $t$ and the undisturbed temperature far away from the sprayed surface $T_{w0}$. For long timescales the material is therefore the limiting factor for the cooling performance of spray cooling.

Rearranging equation~(\ref{QnucleateRem}) leads to
\begin{equation}\label{eq:scale}
\mathcal{T}\equiv\frac{\epsilon_\mathrm{w}^2 \Delta T^2}{\pi {\dot{q}(t)}^2}\approx k(t-t_\mathrm{L}),
\end{equation}
where $k$ is a constant which can be determined from the experiments and $\Delta T = T_\mathrm{w0}-T_\mathrm{sat}$.

\begin{figure}
    \centering
	\includegraphics[width=.5\textwidth]{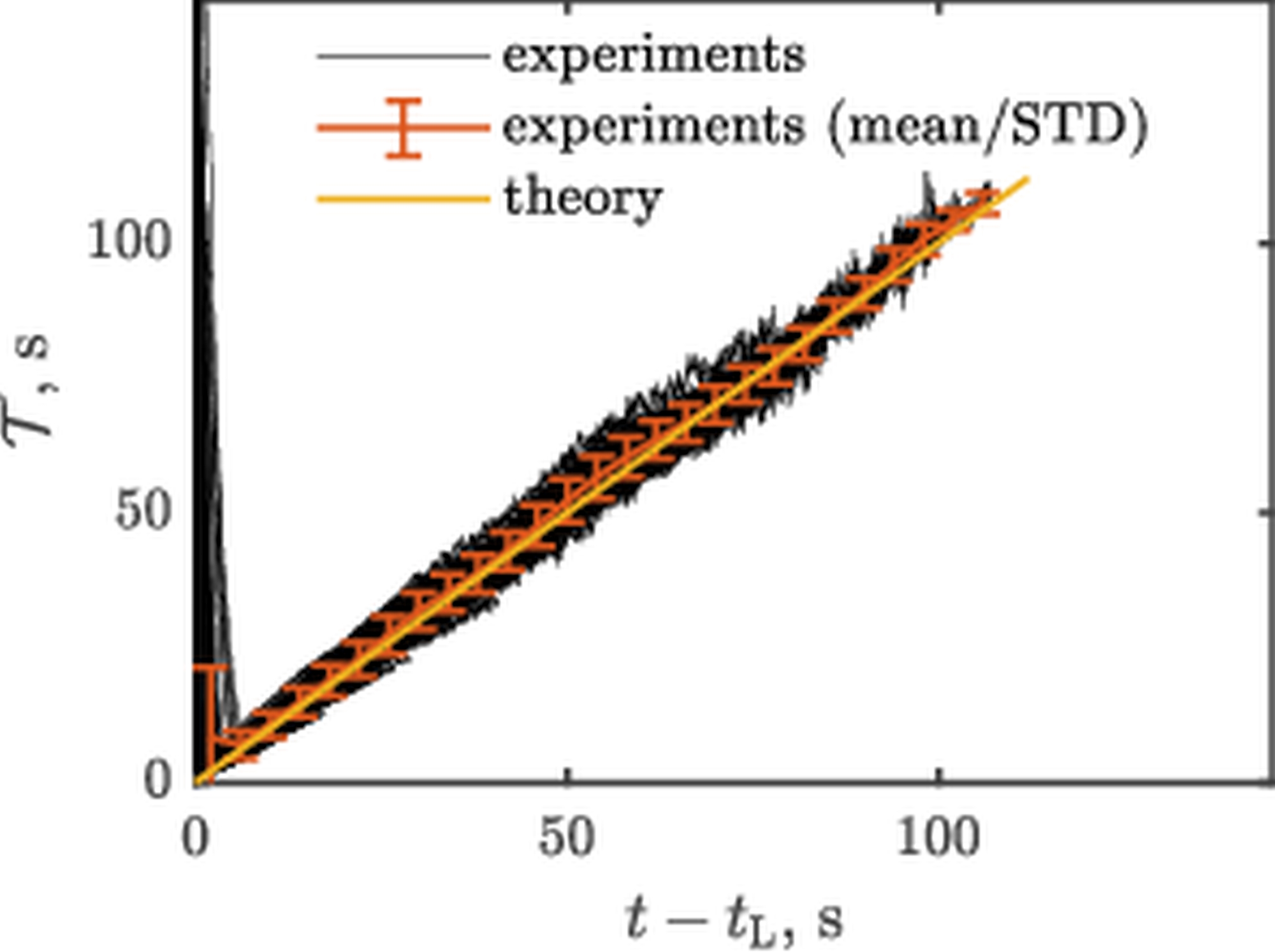}
	\caption{Scaling of the heat flux density, expressed through $\mathcal{T}$, defined in (\ref{eq:scale}),  plotted as a function of time. The straight line, having a slope of unity, indicates good agreement between the theory and the experimental data.}
	\label{fig:nuc_scale}
\end{figure}

In Fig.~\ref{fig:nuc_scale} the values of the term   $\mathcal{T}\equiv\epsilon_\mathrm{w}^2 \Delta T^2/\pi \dot{q}^2$, measured for various spray parameters and different initial substrate temperatures, are shown as a function of $t-t_\mathrm{L}$. The  parameters of the 86 experiments used in the plotting of the data in Fig.~\ref{fig:nuc_scale} are varied over wide ranges: mass flux $\dot{m}$ from 0.5 to 29.5 $\mathrm{kg/m^2s}$, average drop diameter $D_{10}$ from 43 to 78 $\mathrm{\mu m}$, average impact velocity $U$ from 6.7 to 17.7 m/s, initial wall temperature $T_\mathrm{w0}$ from 350 to 450 $\mathrm{^\circ C}$  and spray fluid temperature $T_\mathrm{f0}$ from 18 to 80 $\mathrm{^\circ C}$. Additionally, a line corresponding to the average value of $\mathcal{T}$ for all the experiments and the error bars indicating one standard deviation are shown in the graph. As predicted by the remote asymptotic solution (\ref{eq:scale}), the term $\mathcal{T}$ is very close to the time increment $t-t_\mathrm{L}$ in all the experiments with $k= 1.05$ for all the sets of the operational parameters.

Finally, our experiments show that the heat flux density during fully developed nucleate boiling regime depends significantly on the time. It is estimated as
\begin{equation}\label{QnucleateRemEmpir}
    \dot{q} \approx \frac{\epsilon_\mathrm{w}}{\sqrt{\pi}}\frac{T_\mathrm{w0}-T_\mathrm{sat}}{\sqrt{t-t_\mathrm{L}}}.
\end{equation}

Note that the measured values for $\mathcal{T}$ deviate significantly from the theoretical prediction at small times, associated with the film boiling and transition regimes, for which the scaling (\ref{eq:scale}) is not applicable.

The remote asymptotic solution (\ref{QnucleateRemEmpir})  is valid only for a semi-infinite hot substrate and uniform spray. In practice this means that thickness and the width of the substrate  are much larger than the thickness of the thermal boundary layer in the substrate $\sqrt{\alpha t}$.

\subsection{ Heat flux density $\dot q$ and its upper bound}

 The measured heat flux density during nucleate boiling regime is described very well by the upper bound estimation (\ref{QnucleateRemEmpir}) obtained from the assumption that the interface temperature is equal to $T_\mathrm{sat}$. The reason for this agreement is not immediately clear since some liquid overheat is expected during the nucleate boiling.

 This interesting result can be explained by the fact that the interface temperature oscillates due to the nucleation of multiple bubbles, their expansion and subsequent collapse \cite{Carey}. The bubble  contact lines of the wall-bounded bubbles, whose temperatures are close to $T_\mathrm{sat}$  quickly propagate along the interface. The characteristic time of bubble formation $t_\mathrm{bubble}$ in the experiments \cite{breitenbach2017a} is approximately  $1\, \mathrm{ms}$ . Therefore, the thickness of the thermal boundary layer, associated with a single bubble event is $h_\mathrm{bubble}\sim\sqrt{\alpha t_\mathrm{bubble}}$. This value is approximately 70 $\mu$m in our case. In any case, our measurement system is not able to detect such temperature fluctuations.

 At any given location on the wall surface the temperature jumps to the saturation temperature each time the contact line propagates through this position. Our  measurements allow to estimate only the averaged value of the temperature fluctuations above the saturation temperature. Nevertheless, the contribution of the temperature fluctuations (consisting of positive and negative jumps above $T_\mathrm{sat}$) to the time averaged heat flux $\dot q$ is negligibly small.

\section{Conclusions}
In the experimental part of this study the local heat flux density $\dot q$ and local surface temperature as a function of time were measured during transient spray cooling of a very hot and thick target. The main properties of the spray were accurately characterized in order to better understand their influence on the resulting heat flux density.

The experiments have shown that the  value of $\dot q$ significantly varies during different spray impact regimes: \textit{film boiling regime} at the temperatures above the Leidenfrost point, during which the substrate remains apparently dry; the very short \textit{transition regime} characterized by a rapid increase of the wetted area of the substrate; and \textit{fully developed nucleate boiling regime}, during which the substrate is covered by a continuous thin liquid film created by spray deposition. These regimes are observed in the experiments using a high-speed video system.

A theoretical model for spray cooling is developed which is able to predict the evolution of the temperature profile in a thin thermal boundary layer in the substrate. A remote asymptotic solution for the heat flux density in the fully developed nucleate boiling regime is obtained. The theoretical predictions agree very well with the experimental data over a wide range of spray parameters.

 Compared to existing empirical models the present work represents a different approach, where not only the spray but also the thermal conduction of the substrate and the spray fluid temperature are taken into account. We show that the heat flux in the nucleate boiling regime depends on time and the material. Most studies found in literature deal with  spray cooling with a constant surface temperature. In our case, which is relevant to many practical situations, the process is transient due to the continuous decrease of the substrate temperature. This is why the boiling curves found in many empirical models are not relevant for this kind of processes.

\section{Acknowledgements}
The authors gratefully acknowledge financial support from the Deutsche Forschungsgemeinschaft (DFG) in the framework of SFB-TRR 75 and the Industrieverband Massivumformung e.V.

\bibliographystyle{unsrt}

%\bibliography{references}  %%% Remove comment to use the external .bib file (using bibtex).

\begin{thebibliography}{10}

\bibitem{mudawar2001}
I.~Mudawar.
\newblock {Assessment of high-heat-flux thermal management schemes}.
\newblock {\em IEEE Transactions on Components and Packaging Technologies},
  24(2):122--141, 2001.

\bibitem{Bar-Cohen2006}
Avram Bar-Cohen, Mehmet Arik, and Michael Ohadi.
\newblock {Direct liquid cooling of high flux micro and nano electronic
  components}.
\newblock {\em Proceedings of the IEEE}, 94(8):1549--1570, aug 2006.

\bibitem{Ebadian2011}
M.~A. Ebadian and C.~X. Lin.
\newblock {A Review of High-Heat-Flux Heat Removal Technologies}.
\newblock {\em Journal of Heat Transfer}, 133(11):110801, nov 2011.

\bibitem{Chen1992}
Shih~Jiun Chen and Ampere~A. Tseng.
\newblock {Spray and jet cooling in steel rolling}.
\newblock {\em International Journal of Heat and Fluid Flow}, 13(4):358--369,
  dec 1992.

\bibitem{Pola2013}
A.~Pola, M.~Gelfi, and G.~M. {La Vecchia}.
\newblock {Simulation and validation of spray quenching applied to heavy
  forgings}.
\newblock {\em Journal of Materials Processing Technology}, 213(12):2247--2253,
  dec 2013.

\bibitem{Nizetic2016}
S.~Ni{\v{z}}eti{\'{c}}, D.~{\v{C}}oko, A.~Yadav, and
  F.~Grubi{\v{s}}i{\'{c}}-{\v{C}}abo.
\newblock {Water spray cooling technique applied on a photovoltaic panel: The
  performance response}.
\newblock {\em Energy Conversion and Management}, 108:287--296, jan 2016.

\bibitem{Sargunanathan2016}
S.~Sargunanathan, A.~Elango, and S.~Tharves Mohideen.
\newblock {Performance enhancement of solar photovoltaic cells using effective
  cooling methods: A review}.
\newblock {\em Renewable and Sustainable Energy Reviews}, 64:382--393, oct
  2016.

\bibitem{Chandra1991}
S.~Chandra and C.~T. Avedisian.
\newblock {On the collision of a droplet with a solid surface}.
\newblock {\em Proceedings of the Royal Society A: Mathematical, Physical and
  Engineering Sciences}, 432(1884):13--41, jan 1991.

\bibitem{Bernardin1997b}
John~D. Bernardin, Clinton~J. Stebbins, and Issam Mudawar.
\newblock {Mapping of impact and heat transfer regimes of water drops impinging
  on a polished surface}.
\newblock {\em International Journal of Heat and Mass Transfer},
  40(2):247--267, jan 1997.

\bibitem{Bertola2015a}
V.~Bertola.
\newblock {An impact regime map for water drops impacting on heated surfaces}.
\newblock {\em International Journal of Heat and Mass Transfer}, 85:430--437,
  jun 2015.

\bibitem{Staat2015a}
Hendrik J.J.~J. Staat, Tuan Tran, Bart Geerdink, Guillaume Riboux, Chao Sun,
  Jos{\'{e}}~Manuel Gordillo, and Detlef Lohse.
\newblock {Phase diagram for droplet impact on superheated surfaces}.
\newblock {\em Journal of Fluid Mechanics}, 779:R3, sep 2015.

\bibitem{Roisman2018}
I.~V. Roisman, J.~Breitenbach, and C.~Tropea.
\newblock {Thermal atomisation of a liquid drop after impact onto a hot
  substrate}.
\newblock {\em Journal of Fluid Mechanics}, 842:87--101, may 2018.

\bibitem{Liang2017b}
Gangtao Liang and Issam Mudawar.
\newblock {Review of spray cooling. Part 1: Single-phase and nucleate
  boiling regimes, and critical heat flux}.
\newblock {\em International Journal of Heat and Mass Transfer},
  115(September):1174--1205, dec 2017.

\bibitem{Liang2017d}
Gangtao Liang and Issam Mudawar.
\newblock {Review of spray cooling. Part 2: High temperature boiling regimes
  and quenching applications}.
\newblock {\em International Journal of Heat and Mass Transfer},
  115:1206--1222, 2017.

\bibitem{Cheng2016a}
Wen~Long Cheng, Wei~Wei Zhang, Hua Chen, and Lei Hu.
\newblock {Spray cooling and flash evaporation cooling: The current development
  and application}.
\newblock {\em Renewable and Sustainable Energy Reviews}, 55:614--628, mar
  2016.

\bibitem{Kim2007}
Jungho Kim.
\newblock {Spray cooling heat transfer: The state of the art}.
\newblock {\em International Journal of Heat and Fluid Flow}, 28(4):753--767,
  aug 2007.

\bibitem{Breitenbach2018}
Jan Breitenbach, Ilia~V. Roisman, and Cameron Tropea.
\newblock {From drop impact physics to spray cooling models: a critical
  review}.
\newblock {\em Experiments in Fluids}, 59(3):55, mar 2018.

\bibitem{Mudawar1994}
Issam Mudawar and Thomas~A. Deiters.
\newblock {A universal approach to predicting temperature response of metallic
  parts to spray quenching}.
\newblock {\em International Journal of Heat and Mass Transfer},
  37(3):347--362, feb 1994.

\bibitem{Puschmann2004a}
Frank Puschmann and Eckehard Specht.
\newblock {Transient measurement of heat transfer in metal quenching with
  atomized sprays}.
\newblock {\em Experimental Thermal and Fluid Science}, 28(6):607--615, jun
  2004.

\bibitem{Wendelstorf2008}
J.~Wendelstorf, K.~H. Spitzer, and R.~Wendelstorf.
\newblock {Spray water cooling heat transfer at high temperatures and liquid
  mass fluxes}.
\newblock {\em International Journal of Heat and Mass Transfer},
  51(19-20):4902--4910, sep 2008.

\bibitem{Yang1996}
J.~Yang, L.~C. Chow, and M.~R. Pais.
\newblock {Nucleate Boiling Heat Transfer in Spray Cooling}.
\newblock {\em Journal of Heat Transfer}, 118(3):668, aug 1996.

\bibitem{Chen2002a}
Ruey-Hung~Hung Chen, Louis~C Chow, and Jose~E Navedo.
\newblock {Effects of spray characteristics on critical heat flux in subcooled
  water spray cooling}.
\newblock {\em International Journal of Heat and Mass Transfer},
  45(19):4033--4043, sep 2002.

\bibitem{Estes1995a}
Kurt~A. Estes and Issam Mudawar.
\newblock {Correlation of sauter mean diameter and critical heat flux for spray
  cooling of small surfaces}.
\newblock {\em International Journal of Heat and Mass Transfer},
  38(16):2985--2996, nov 1995.

\bibitem{Cebo-Rudnicka2016}
Agnieszka Cebo-Rudnicka, Zbigniew Malinowski, and Andrzej Buczek.
\newblock {The influence of selected parameters of spray cooling and thermal
  conductivity on heat transfer coefficient}.
\newblock {\em International Journal of Thermal Sciences}, 110:52--64, dec
  2016.

\bibitem{Woodfield2006}
P.~L. Woodfield, M.~Monde, and Y.~Mitsutake.
\newblock {Improved analytical solution for inverse heat conduction problems on
  thermally thick and semi-infinite solids}.
\newblock {\em International Journal of Heat and Mass Transfer},
  49(17-18):2864--2876, 2006.

\bibitem{Lefebvre2018}
Arthur Lefebvre.
\newblock {\em {Atomization and Sprays}}.
\newblock CRC Press, Boca Raton, dec 1988.

\bibitem{Ozsk1980}
M.~Necati {\"O}z{\i}{\c{s}}{\i}k.
\newblock {\em {Heat conduction}}.
\newblock John Wiley {\&} Sons, New York, Chichester, Brisbane, Toronto, 1980.

\bibitem{Breitenbach2017}
Jan Breitenbach, Ilia~V. Roisman, and Cameron Tropea.
\newblock {Heat transfer in the film boiling regime: Single drop impact and
  spray cooling}.
\newblock {\em International Journal of Heat and Mass Transfer}, 110:34--42,
  jul 2017.

\bibitem{Tran2012}
Tuan Tran, J~J Staat, Hendrik, Andrea Prosperetti, Chao Sun, and Detlef Lohse.
\newblock Drop impact on superheated surfaces.
\newblock {\em Physical Review Letters}, 108(3):1--5, 2012.

\bibitem{Chaze2019}
W.~Chaze, O.~Caballina, G.~Castanet, J.~F. Pierson, F.~Lemoine, and D.~Maillet.
\newblock {Heat flux reconstruction by inversion of experimental infrared
  temperature measurements. Application to the impact of a droplet in the
  film boiling regime}.
\newblock {\em International Journal of Heat and Mass Transfer}, 128:469--478,
  jan 2019.

\bibitem{breitenbach2017a}
Jan Breitenbach, Ilia~V. Roisman, and Cameron Tropea.
\newblock {Drop collision with a hot, dry solid substrate: Heat transfer during
  nucleate boiling}.
\newblock {\em Physical Review Fluids}, 2(7):074301, jul 2017.

\bibitem{Abu-Zaid2004}
M.~Abu-Zaid.
\newblock {An experimental study of the evaporation characteristics of
  emulsified liquid droplets}.
\newblock {\em Heat and Mass Transfer}, 40(9):737--741, jun 2003.

\bibitem{Itaru1978}
Michiyoshi Itaru and Makino Kunihide.
\newblock {Heat transfer characteristics of evaporation of a liquid droplet on
  heated surfaces}.
\newblock {\em International Journal of Heat and Mass Transfer},
  21(5):605--613, may 1978.

\bibitem{Tartarini1999}
P.~Tartarini, G.~Lorenzini, and M.~R. Randi.
\newblock {Experimental study of water droplet boiling on hot, non-porous
  surfaces}.
\newblock {\em Heat and Mass Transfer}, 34(6):437--447, apr 1999.

\bibitem{Carey}
Van~P. Carey.
\newblock {\em {Liquid Vapor Phase Change Phenomena: An Introduction to the
  Thermophysics of Vaporization and Condensation Processes in Heat Transfer
  Equipment}}.
\newblock Taylor and Francis, New York, 2. ed edition, 2018.

\end{thebibliography}
%%% and comment out the ``thebibliography'' section.

\end{document}